\documentclass[prb,twocolumn,superscriptaddress,showpacs,notitlepage,nofootinbib]{revtex4-1}
\usepackage{amsfonts}
\usepackage{amsmath}
\usepackage{amsthm}
\usepackage{braket}
\usepackage{graphicx}
\usepackage{subcaption}
\usepackage{hyperref}
\usepackage[dvipsnames]{xcolor}
\usepackage{amssymb}
\usepackage{dsfont}
\usepackage{verbatim}
\usepackage{amsmath,amscd}
\usepackage[all,cmtip]{xy}
\usepackage{multirow}

\definecolor{darkblue}{RGB}{0,0,127} 
\definecolor{darkgreen}{RGB}{0,130,80}
\definecolor{darkred}{RGB}{160,20,20}
\hypersetup{
	colorlinks,
	linkcolor=darkblue,
	citecolor=darkgreen,
	filecolor=red,
	urlcolor=blue,
	pdftitle={},
	pdfauthor={Dominic J. Williamson}
}

\newcommand{\red}[1]{\textcolor{darkred}{#1}}

\graphicspath{{./Figures/}}

\usepackage{tikz,pgfplots}\pgfplotsset{compat=newest}
\usetikzlibrary{external,calc,decorations.pathreplacing,decorations.markings,decorations.pathmorphing,arrows.meta,shapes.geometric}
\newlength\figureheight
\newlength\figurewidth

\tikzset{
    double color fill/.code 2 args={
        \pgfdeclareverticalshading[%
            tikz@axis@top,tikz@axis@middle,tikz@axis@bottom%
        ]{diagonalfill}{100bp}{%
            color(0bp)=(tikz@axis@bottom);
            color(50bp)=(tikz@axis@bottom);
            color(50bp)=(tikz@axis@middle);
            color(50bp)=(tikz@axis@top);
            color(100bp)=(tikz@axis@top)
        }
        \tikzset{shade, left color=#1, right color=#2, shading=diagonalfill}
    }
}

\newcommand{\includeTikz}[2]{
\includegraphics{#1}
}

\newcommand{\Z}{\ensuremath{\mathbb{Z}}}

\newcommand{\drawgenerator}[8]{%
\xymatrix@!0{%
& #8 \ar@{-}[ld]\ar@{.}[dd] \ar@{-}[rr] & & #7 \ar@{-}[ld]  \\%
#1 \ar@{-}[rr] \ar@{-}[dd] &  & #2 \ar@{-}[dd] &            \\%
& #6 \ar@{.}[ld] &  & #5 \ar@{-}[uu] \ar@{.}[ll]       \\%
#3 \ar@{-}[rr] &  & #4 \ar@{-}[ru]                       %
}%
}

\newcommand{\plaquette}[4]{
\xymatrix@!0{%
#1 \ar@{-}[r] \ar@{-}[d]  & #2 \ar@{-}[d] 
\\
#3 \ar@{-}[r]  & #4
}
}
\newcommand{\plaquettedoublespacing}[4]{
\xymatrix@!0{%
#1 \ar@{-}[rr] \ar@{-}[d]  & & #2 \ar@{-}[d] 
\\
#3 \ar@{-}[rr]  & & #4
}
}
\newcommand{\plaquettetriplespacing}[4]{
\xymatrix@!0{%
#1 \ar@{-}[rrrr] \ar@{-}[d] &  & & & #2 \ar@{-}[d] 
\\
#3 \ar@{-}[rrrr]  & & & & #4
}
}

\newcommand{\qd}{\ensuremath{\mathcal{D}}}
\newcommand{\tr}{\ensuremath{\text{Tr}}}

\begin{document}

  \title{Spurious topological entanglement entropy from subsystem symmetries}  
  \author{Dominic~J. Williamson}
  \affiliation{Department of Physics, Yale University, New Haven, CT 06520-8120, USA}
  \author{Arpit Dua}
  \affiliation{Department of Physics, Yale University, New Haven, CT 06520-8120, USA}
  \affiliation{Yale Quantum Institute, Yale University, New Haven, CT 06520, USA}
  \author{Meng Cheng}
  \affiliation{Department of Physics, Yale University, New Haven, CT 06520-8120, USA}

  \begin{abstract}
We demonstrate that linear combinations of subregion entropies with canceling boundary terms, commonly used to calculate the topological entanglement entropy, may suffer from spurious nontopological contributions even in models with zero correlation length. 
These spurious contributions are due to a specific kind of long-range string order, and persist throughout certain subsystem symmetry-protected phases. We introduce an entropic quantity that measures the presence of such order, and hence should serve as an order parameter for the aforementioned phases. 
  \end{abstract}

  \maketitle

  \noindent\textbf{ Introduction.}
  \label{sec:intro}
  Topologically ordered phases are characterized by non-local quantum entanglement in groundstate wavefunctions~\cite{KitaevPreskill,levinwenentanglement,chen2010local,Bultinck2017183}. One of the simplest and most widely used diagnostics for topological order is the constant correction to the area law scaling of the entanglement entropy~\cite{KitaevPreskill,levinwenentanglement}, known as the \emph{topological entanglement entropy} (TEE). More precisely, in ground states of local gapped spin systems in two dimensions, the entanglement entropy of an arbitrary region $R$ with smooth boundary $\partial R$ is expected to scale as~\cite{KitaevPreskill} 
\begin{align}
\label{eq:arealaw}
S_R = c |\partial R |  - \gamma + \dots
\, ,
\end{align}
for some nonuniversal constant $c$. The TEE $-\gamma$ is a nonpositive constant correction to the area law. Refs.~\onlinecite{KitaevPreskill,levinwenentanglement} argued that $\gamma=\log \qd$, using ideas based on the topological quantum field theories that describe 2D gapped phases in the long-wavelength limit. 
The quantity $\qd$ is the total quantum dimension of superselection sectors in a model. 
TEE is related to the so-called boundary entropy in $(1+1)$D quantum critical theories~\cite{Affleck1991, Fendley2007}.

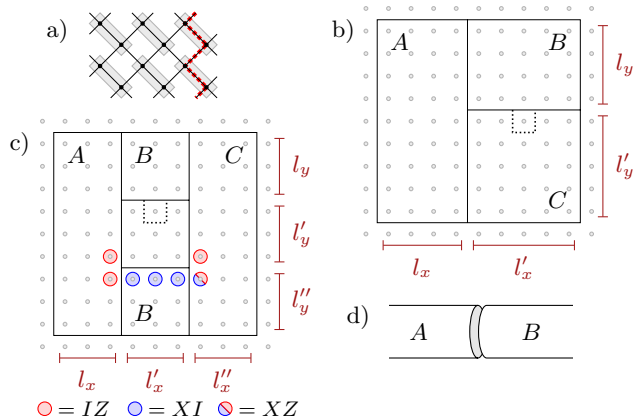
\begin{figure}[t]
\center
	\includeTikz{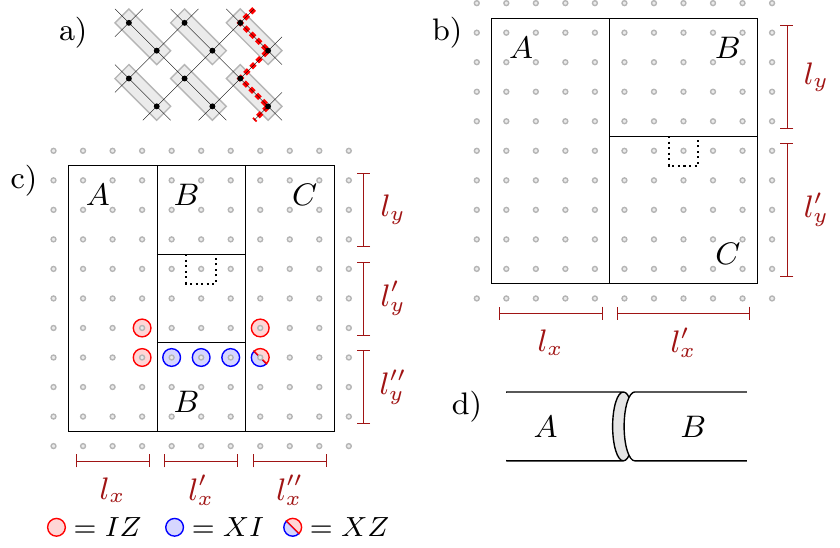}{
	\begin{tikzpicture}[scale=.1]
	\def\dx{4};
	\def\l{5};
	\def\hl{2}; 
		\def\hlm{2}; 
	\def\r{.3};
	\begin{scope}[xshift=2cm,yshift=20cm,rotate=-45]
\node[transform shape,draw=black!29,fill=black!10!white!80,minimum height=.5*\dx cm,minimum width=1.5*\dx cm] at (2*1*\dx+.5*\dx,0*\dx){};
\node[transform shape,draw=black!29,fill=black!10!white!80,minimum height=.5*\dx cm,minimum width=1.5*\dx cm] at (2*1*\dx+.5*\dx,2*\dx){};
\node[transform shape,draw=black!29,fill=black!10!white!80,minimum height=.5*\dx cm,minimum width=1.5*\dx cm] at (2*2*\dx+.5*\dx,2*\dx){};
\node[transform shape,draw=black!29,fill=black!10!white!80,minimum height=.5*\dx cm,minimum width=1.5*\dx cm] at (2*1*\dx-.5*\dx,1*\dx){};
\node[transform shape,draw=black!29,fill=black!10!white!80,minimum height=.5*\dx cm,minimum width=1.5*\dx cm] at (2*2*\dx-.5*\dx,1*\dx){};
\node[transform shape,draw=black!29,fill=black!10!white!80,minimum height=.5*\dx cm,minimum width=1.5*\dx cm] at (2*2*\dx-.5*\dx,3*\dx){};
\draw[draw=red,line width=1.5pt,densely dotted]  (3*\dx,3.5*\dx) -- (3*\dx,3*\dx) -- (4*\dx,3*\dx) -- (4*\dx,2*\dx) -- (5*\dx,2*\dx) -- (5*\dx,1.5*\dx) ;
\draw[draw=black,line width=.1pt] (1.5*\dx,0*\dx) -- (3.5*\dx,0*\dx) ;
\draw[draw=black,line width=.1pt] (0.5*\dx,1*\dx) -- (4.5*\dx,1*\dx) ;
\draw[draw=black,line width=.1pt] (1.5*\dx,2*\dx) -- (5.5*\dx,2*\dx) ;
\draw[draw=black,line width=.1pt] (2.5*\dx,3*\dx) -- (4.5*\dx,3*\dx) ;
\draw[draw=black,line width=.1pt] (1*\dx,.5*\dx) -- (1*\dx,1.5*\dx) ;
\draw[draw=black,line width=.1pt] (2*\dx,-.5*\dx) -- (2*\dx,2.5*\dx) ;
\draw[draw=black,line width=.1pt] (3*\dx,-.5*\dx) -- (3*\dx,3.5*\dx) ;
\draw[draw=black,line width=.1pt] (4*\dx,.5*\dx) -- (4*\dx,3.5*\dx) ;
\draw[draw=black,line width=.1pt] (5*\dx,1.5*\dx) -- (5*\dx,2.5*\dx) ;
  \fill[color=black] (2*\dx,2*\dx) circle (\r);  \fill[color=black] (3*\dx,2*\dx) circle (\r);   \fill[color=black] (4*\dx,2*\dx) circle (\r);  \fill[color=black] (5*\dx,2*\dx) circle (\r);
 \fill[color=black] (1*\dx,1*\dx) circle (\r);  \fill[color=black] (2*\dx,1*\dx) circle (\r);  \fill[color=black] (3*\dx,1*\dx) circle (\r);  \fill[color=black] (4*\dx,1*\dx) circle (\r);
  \fill[color=black] (2*\dx,0*\dx) circle (\r);  \fill[color=black] (3*\dx,0*\dx) circle (\r);
\fill[color=black] (3*\dx,3*\dx) circle (\r); \fill[color=black] (4*\dx,3*\dx) circle (\r);
	\end{scope}
\def\dy{3}
\def\w{.25}
\def\n{10}
\def\de{.25}
	\begin{scope}[xshift=43cm,yshift=-8cm]
			\foreach \x in {0,1,...,\n} {
        \foreach \y in {0,1,...,\n} {
		  \draw[color=black!29,fill=black!10!white!80] (\x*\dy,\y*\dy) circle (\w);
        }
        }
	\draw[draw=black,line width=.2pt] (.5*\dy,.5*\dy) -- (9.5*\dy,.5*\dy) --  (9.5*\dy,9.5*\dy) -- (.5*\dy,9.5*\dy) -- cycle 
	 (4.5*\dy,.5*\dy) --  (4.5*\dy,9.5*\dy) 
 	 (4.5*\dy,5.5*\dy) --  (9.5*\dy,5.5*\dy) ;
 	 \draw[draw=black,line width=.6pt,densely dotted]  	 (6.5*\dy,5.5*\dy) -- (6.5*\dy,4.5*\dy) -- (7.5*\dy,4.5*\dy) --  (7.5*\dy,5.5*\dy) ;
   	\draw[draw=darkred,line width=.2pt,|-|]    (4*\dy+\de*\dy,-.5*\dy) to (1*\dy -\de*\dy, -.5*\dy);
   	\draw[draw=darkred,line width=.2pt,|-|]    (5*\dy-\de*\dy,-.5*\dy) to (9*\dy +\de*\dy, -.5*\dy);
   	\draw[draw=darkred,line width=.2pt,|-|]    (10.5*\dy , 5*\dy+\de*\dy) to (10.5*\dy , 1*\dy-\de*\dy);
   	\draw[draw=darkred,line width=.2pt,|-|]    (10.5*\dy , 6*\dy-\de*\dy) to (10.5*\dy , 9*\dy+\de*\dy);
	\node at (2.5*\dy,-1.5*\dy){\red{$l_x$}};
	\node at (7*\dy,-1.5*\dy){\red{$l_x'$}};
	\node at (11.5*\dy,3*\dy){\red{$l_y'$}};
	\node at (11.5*\dy,7.5*\dy){\red{$l_y$}};
	\node at (1.5*\dy,8.5*\dy){$A$};
	\node at (8.5*\dy,8.5*\dy){$B$};
	\node at (8.5*\dy,1.5*\dy){$C$};
	\end{scope}
	\begin{scope}[xshift=0cm,yshift=-23cm]
			\foreach \x in {4,5,6} {
			\node[transform shape,draw=blue,fill=blue!80!white!20,circle,minimum size=1.8cm] at (\x*\dy,3*\dy) {};
        }	
			\node[transform shape,draw=red,fill=red!80!white!20,circle,minimum size=1.8cm] at (3*\dy,3*\dy) {};
			\node[transform shape,draw=red,fill=red!80!white!20,circle,minimum size=1.8cm] at (3*\dy,4*\dy) {};
			\node[transform shape,draw=blue,semicircle,minimum size=.9cm,anchor=south,rotate=135,fill={blue!80!white!20}] at (7*\dy,3*\dy) {};
			\node[transform shape,draw=red,semicircle,minimum size=.9cm,anchor=south,rotate=-45,fill={red!80!white!20}] at (7*\dy,3*\dy) {};
			\node[transform shape,draw=red,fill=red!80!white!20,circle,minimum size=1.8cm] at (7*\dy,4*\dy) {};
			\foreach \x in {0,1,...,\n} {
        \foreach \y in {0,1,...,\n} {
		  \draw[color=black!29,fill=black!10!white!80] (\x*\dy,\y*\dy) circle (\w);
        }
        }
	\draw[draw=black,line width=.2pt] (.5*\dy,.5*\dy) -- (9.5*\dy,.5*\dy) --  (9.5*\dy,9.5*\dy) -- (.5*\dy,9.5*\dy) -- cycle 
	 (3.5*\dy,.5*\dy) --  (3.5*\dy,9.5*\dy) 	 (6.5*\dy,.5*\dy) --  (6.5*\dy,9.5*\dy) 
 	 (3.5*\dy,3.5*\dy) --  (6.5*\dy,3.5*\dy)  	 (3.5*\dy,6.5*\dy) --  (6.5*\dy,6.5*\dy)  ;
 	 \draw[draw=black,line width=.6pt,densely dotted]  	 (4.5*\dy,6.5*\dy) -- (4.5*\dy,5.5*\dy) -- (5.5*\dy,5.5*\dy) --  (5.5*\dy,6.5*\dy) ;
   	\draw[draw=darkred,line width=.2pt,|-|]    (3*\dy+\de*\dy,-.5*\dy) to (1*\dy -\de*\dy, -.5*\dy);
   	\draw[draw=darkred,line width=.2pt,|-|]    (4*\dy-\de*\dy,-.5*\dy) to (6*\dy +\de*\dy, -.5*\dy);
   	\draw[draw=darkred,line width=.2pt,|-|]    (9*\dy+\de*\dy,-.5*\dy) to (7*\dy -\de*\dy, -.5*\dy);
   	\draw[draw=darkred,line width=.2pt,|-|]    (10.5*\dy , 3*\dy+\de*\dy) to (10.5*\dy , 1*\dy-\de*\dy);
   	\draw[draw=darkred,line width=.2pt,|-|]    (10.5*\dy , 4*\dy-\de*\dy) to (10.5*\dy , 6*\dy+\de*\dy);
    \draw[draw=darkred,line width=.2pt,|-|]    (10.5*\dy , 7*\dy-\de*\dy) to (10.5*\dy , 9*\dy+\de*\dy);
	\node at (2*\dy,-1.5*\dy){\red{$l_x$}};
	\node at (5*\dy,-1.5*\dy){\red{$l_x'$}};
	\node at (8*\dy,-1.5*\dy){\red{$l_x''$}};
	\node at (11.5*\dy,2*\dy){\red{$l_y''$}};
	\node at (11.5*\dy,5*\dy){\red{$l_y'$}};
	\node at (11.5*\dy,8*\dy){\red{$l_y$}};
	\node at (1.5*\dy,8.5*\dy){$A$};
	\node at (4.5*\dy,8.5*\dy){$B$};
	\node at (4.5*\dy,1.5*\dy){$B$};
	\node at (8.5*\dy,8.5*\dy){$C$};
	\node[transform shape,draw=red,fill=red!80!white!20,circle,minimum size=1.8cm] at (.1*\dy,-2.75*\dy) {};
	\node[anchor=west] at (.1*\dy+.2*\dy,-2.75*\dy) {\footnotesize $=IZ$};
	\node[transform shape,draw=blue,fill=blue!80!white!20,circle,minimum size=1.8cm] at (4.1*\dy,-2.75*\dy) {};
	\node[anchor=west] at (4.1*\dy+.2*\dy,-2.75*\dy) {\footnotesize $=XI$};
			\node[transform shape,draw=blue,semicircle,minimum size=.9cm,anchor=south,rotate=135,fill={blue!80!white!20}] at  (8.1*\dy,-2.75*\dy) {};
			\node[transform shape,draw=red,semicircle,minimum size=.9cm,anchor=south,rotate=-45, fill={red!80!white!20}] at (8.1*\dy,-2.75*\dy) {};
	\node[anchor=west] at (8.1*\dy+.2*\dy,-2.75*\dy) {\footnotesize $=XZ$};
	\end{scope}
	\begin{scope}[xshift=51cm,yshift=-21cm]
	\node at (0,0) [cylinder,  draw,minimum height=1.4cm,minimum width=.7cm, cylinder uses custom fill, cylinder end fill=black!10] {};
		\node at (13.5,0) [cylinder,  draw,minimum height=1.5cm,minimum width=.7cm, cylinder uses custom fill, cylinder end fill=black!10] {};
		\node at (-1,0){$A$};
	\node at (14,0){$B$};
	\fill[color=white] (-5,-4) -- (-7,-4) -- (-7,4) -- (-5,4) -- cycle;
	\fill[color=white] (19.5,-4) -- (22.5,-4) -- (22.5,4) -- (19.5,4) -- cycle;
	\end{scope}
	\node at (2cm,19cm){a)};
	\node at (40cm,19cm){b)};
	\node at (-3cm,4cm){c)};
	\node at (42cm,-19cm){d)};
	\end{tikzpicture}
	}
 \caption{a) Coarse grained lattice used for the cluster state. b) The regions used to calculate $S^{\text{KP}}_{\text{topo}}$. c) The regions used to calculate $S^{\text{LW}}_{\text{topo}}$ and a spurious nonlocal stabilizer for Hamiltonian~\eqref{eq:cstate}. d) Bipartition of an infinite cylinder. }
\label{clusterlattice}
\end{figure}

To calculate the TEE in a given lattice model, one must separate the constant correction to the entanglement entropy from nonuniversal contributions due to boundaries, corners and other lattice details. 
This necessarily requires  one to consider regions much larger than the correlation length of a model. 
 A numerically convenient method to extract the TEE is to consider the vacuum minimally entangled state on an infinite cylinder, where a bipartition across a cut with no corners can easily be made, see Fig.~\ref{clusterlattice}~d). The TEE term can then be extracted by a scaling analysis of the half cylinder entanglement entropy with the radius of the cylinder~\cite{jiang2012identifying}. In Refs.~\onlinecite{bravyiunpublished,PhysRevB.92.075104,PhysRevB.94.075151,santos2018symmetry} it was demonstrated that this method potentially suffers spurious contributions due to nontrivial symmetry-protected states along the cut. 
A quantity named the replica correlation length was identified in Ref.~\onlinecite{PhysRevB.94.075151}, and the authors conjectured that both conventional and replica correlation lengths must be relatively small for the TEE to be reliably extracted via cylinder extrapolation. 
 
The conceptually cleanest way to extract the TEE is to take a cleverly defined linear combination of subregion entropies, due to Kitaev and Preskill~\cite{KitaevPreskill}, and Levin and Wen~\cite{levinwenentanglement}, that is designed to cancel boundary contributions and furthermore corner contributions which are unavoidable in a lattice model: 
\begin{equation}
	 \begin{split}
 S_{\text{topo}} &= I_3(A,B,C) 
 \\
&= S_A + S_B + S_C - S_{AB} - S_{BC} - S_{AC} 
 + S_{ABC} 
 \nonumber
\, .
\end{split}
	\label{}
\end{equation}
Here $I_3(A,B,C)$ is the tripartite information of three regions $A,B,C$. For the choice of regions shown in Fig.~\ref{clusterlattice}~b) Ref.~\onlinecite{KitaevPreskill} argued that ${S_{\text{topo}}^{\text{KP}} \approx -\gamma}$ while for the regions shown in Fig.~\ref{clusterlattice}~c) both Refs.~\onlinecite{KitaevPreskill,levinwenentanglement} argued that ${S^{\text{LW}}_{\text{topo}} \approx - 2 \gamma}$, where KP and LW denote the different choice of regions respectively. It was argued that the preceding approximate equalities should become exact as the sizes of the regions become much larger than the correlation length.

 A crucial ingredient in the argument that $S_{\text{topo}}$ measures a topological quantity is its invariance under deformations of the regions used to calculate it. In particular, it is required that under small deformations of the $B$ region, far from region $A$, the following changes of subregion entropies cancel $\Delta S_{AB} - \Delta S_A \approx 0$, and similarly for deformations of other regions. An analogous condition is required for deformations of a triple point where $A$, $B$ and $C$ meet. 
 The authors of Ref.~\onlinecite{PhysRevB.94.075151} did not find any cases where this method suffered from spurious contributions when applied to translationally invariant models\footnote{There is a very inhomogenous example due to Bravyi that does suffer a spurious contribution~\cite{bravyiunpublished,PhysRevB.94.075151}.}.

In this letter we demonstrate that the Kitaev-Preskill and Levin-Wen schemes to calculate the TEE may, in fact, suffer from spurious contributions when applied to translationally invariant models. 
Specifically, we find translationally invariant Hamiltonians with zero correlation length where the TEE extracted from a linear combination of subregion entropies differs from the true value by a nonzero integer, for arbitrarily large regions. 
This occurs for regions whose boundaries run along lattice directions that have an associated long-range string order. 
Such long-range string order is characteristic of an entire nontrivial \emph{subsystem symmetry-protected topological} (SSPT) phase~\cite{
PhysRevLett.103.020506,PhysRevB,raussendorf2018computationally,you2018subsystem,devakul2018fractal,kubica2018ungauging,you2018symmetric,devakul2018universal,stephen2018subsystem,subsystemphaserel}. We introduce an entropic quantity, that is sensitive to the shape but not size of the region for which it is defined, which can robustly detect the presence of spurious contributions to the TEE.

We focus our attention on zero correlation length Pauli stabilizer Hamiltonians to avoid finite-size effects. The Hamiltonians are defined on the square lattice with $Q$ qubits per site and  $Q$ independent stabilizer generators per plaquette. We assume there are no relations between the stabilizers on disc-like patches (otherwise the ground space would be under-constrained and possess local logical operators). In the case that each stabilizer acts nontrivially on all sites adjacent to its plaquette, we find $S^{\text{KP}}_{\text{topo}}=-Q$ for the regions drawn in Fig.~\ref{clusterlattice}~b). We demonstrate below that this does not always match the expected $\gamma$, and may be sensitive to the shape of the regions, in violation of assumptions that are relied upon to argue $S^{\text{KP}}_{\text{topo}}\approx-\gamma$. 
\newline

 \noindent \textbf{ An example with spurious TEE.}
 \label{sec:example}
The main example we consider is simply a \emph{cluster state}~\cite{PhysRevLett.86.5188} Hamiltonian, with qubits on the sites of a square lattice. Recently it has been recognized that this cluster state lies in a nontrivial \emph{strong} SSPT phase~\cite{you2018subsystem,subsystemphaserel}, which is a gapped quantum phase of matter protected by rigid string symmetries that is not equivalent to decoupled 1D chains. 
 Our definition of the cluster state differs from the usual one by some coarse graining as shown in Fig.~\ref{clusterlattice}~a), such that the rigid string symmetries in Eq.~\eqref{eq:subgen} align with the lattice directions. There are two qubits per site and the local stabilizer generators are given by translates of 
 \begin{align}
 \label{eq:cstate}
 \begin{array}{c}
 \plaquette{IZ}{IZ}{IZ}{XZ}
 \quad
 \plaquette{ZX}{ZI}{ZI}{ZI}
 \end{array}
 \, .
\end{align}
Here $X$/$Z$ stand for Pauli X/Z matrices, and $I$ is the identity matrix. For completeness we have included relevant background material about stabilizer Hamiltonians, and our notation for quantum gates, in the appendix.

The ground state of Hamiltonian~\eqref{eq:cstate} can be constructed by applying a local unitary circuit, consisting of $CZ$ gates applied to pairs of qubits connected by edges of the original square lattice, to the $\ket{+}^{\otimes N}$ state, where $\ket{+}$ is the $+1$ eigenstate of $X$. Hence the Hamiltonian lies in the trivial topological phase and  $\mathcal{D} = 1$. Furthermore, since the correlation length is zero one would also expect $S_{\text{topo}}=0$. However, this is not the case. For the regions shown in Fig.~\ref{clusterlattice}~b) we find $S^{\text{KP}}_{\text{topo}}=-2$. 
This follows from the general consideration in the previous section, as there are no relations between the stabilizer generators and no nonlocal stabilizers for the regions considered. For the regions shown in Fig.~\ref{clusterlattice}~c) we find $S^{\text{LW}}_{\text{topo}}=-2$, and from the cylinder extrapolation method using the regions shown in  Fig.~\ref{clusterlattice}~d) we find a $-1$ correction to the area law. 

In fact, we find that both $S_{\text{topo}}$ quantities are not ``topological invariants'': first, they may jump by an integer under small deformations of the regions, in particular $S^{\text{KP}}_{\text{topo}}=-1$ after deforming the boundary to the dotted line in Fig.~\ref{clusterlattice}~b). This lack of deformation invariance is due to a difference in the change of entropies ${\Delta S_{B} - \Delta S_{AB} = 1}$, which violates the conditions required for  $S^{\text{KP}}_{\text{topo}}$ to be insensitive to the choice of regions. 
Similarly, we find $S^{\text{LW}}_{\text{topo}}=-1$ after a small deformation of the boundary to the dotted line in Fig.~\ref{clusterlattice}~c). 
A similar property holds for the cylinder extrapolation method, as was previously observed in Ref.~\onlinecite{PhysRevB.94.075151}. 
Secondly, both $S_{\text{topo}}$ quantities may jump by an integer when a local unitary circuit is applied. Specifically, applying a product of ${(H\otimes H) CZ  (H\otimes H)} $ gates to pairs of qubits that are nearest neighbors along vertical lines through the lattice, given by translates of the dashed red line in Fig.~\ref{clusterlattice}~a), we find ${S^{\text{KP}}_{\text{topo}}=S^{\text{LW}}_{\text{topo}}=-4}$. The definition of the above gates is given in the appendix.

To the best of our knowledge these spurious values of $S_{\text{topo}}$ for the square lattice cluster state were not previously noticed. This is likely because the majority of generic regions (not running along the directions of rigid symmetry) lead to $S_{\text{topo}}=0$~\cite{PhysRevB.97.134426}. Furthermore, although ${\Delta S_{AB} - \Delta S_B \neq 0}$, for a large range of deformations  $S_{\text{topo}}$ may remain invariant due to cancellations such as
${\Delta S_{AB} - \Delta S_B  + \Delta S_{AC} - \Delta S_C = 0}$. 
\newline

\noindent\textbf{Rigid string symmetries.}
We now demonstrate that the spurious contribution to $S_{\text{topo}}$, and its jump under both local unitaries and small deformations of the regions from which it is calculated, can be accounted for by long-range string order.

Suppose the model of interest has a string-like symmetry~\cite{PhysRevLett.86.5188,PhysRevLett.103.020506,PhysRevB,raussendorf2018computationally,you2018subsystem,devakul2018fractal,kubica2018ungauging,you2018symmetric,devakul2018universal,stephen2018subsystem,subsystemphaserel}, which have previously been studied under the name of gauge-like symmetry~\cite{Batista2005-yr,Nussinov06102009,NUSSINOV2009977}, and consider the symmetry action restricted to an open segment of string. If the stringlike symmetry is not spontaneously broken, it cannot create topologically nontrivial particles at the endpoints. Hence it should be possible to locally annihilate the excitations created at the end points by acting in a neighborhood that contains them. The combined action of the open string followed by endpoint annihilation operators preserves the ground space. 
In particular for a stabilizer model that satisfies the topological order condition~\cite{bravyi2010topological,haah2013lattice,haah2013commuting} the product of the open stringlike operator with the endpoint annihilation operators is again a stabilizer. 

For the cluster example in Eq.~\eqref{eq:cstate}, the relevant subsystem symmetry~\cite{subsystemphaserel} is given by a representation of $\Z_2 \times \Z_2$ along each horizontal and vertical line generated by
\begin{align}
\label{eq:subgen}
\bigotimes_i (IX)_{i,j}
\, ,
\ 
\bigotimes_i (XI)_{i,j}
\, ,
\ 
\bigotimes_j (IX)_{i,j}
\, , 
\
\bigotimes_j (XI)_{i,j}
\, .
\end{align}
The end point operators for the horizontal $XI$ and $IX$ symmetries in Eq.~\eqref{eq:subgen} are given by 
 \begin{align}
\label{eq:csendpointops}
 \begin{array}{c}
 \xymatrix@!0{%
IZ\ar@{-}[d]  
\\
IZ 
}
\end{array}
\, ,
&&
  \begin{array}{c}
 \xymatrix@!0{%
ZI\ar@{-}[d]  
\\
ZI 
}
 \end{array}
 \, ,
\end{align}
respectively. Similarly, for the symmetries along the vertical direction the endpoint operators are given by rotating the above by $90$ degrees.

If the stringlike symmetries are products of generators that have a \emph{wider} support than the string itself, they may appear as nonlocal stabilizers at boundaries that run along the direction of the subsystem symmetry, such as for a bipartition of a cylinder. 
In this case the endpoint operators must also have a wider support than the string. Hence for a region contained within a disc, 
nonlocal stabilizers made up of symmetry strings with end point operators should only occur at boundaries along the direction of the subsystem symmetry with protrusions that accommodate each end point operator, such as the dotted deformations shown in Fig.~\ref{clusterlattice}~b),~c). 

The appearance of nonlocal stabilizers is an important qualitative distinction between unbroken rigid stringlike symmetries and the more familiar deformable 1-form symmetries~\cite{devakul2018universal}. For unbroken 1-form symmetries, a stabilizer given by the product of a symmetry string with endpoint operators cannot be nonlocal for a sufficiently large region as it can be deformed into the bulk of the region and then decomposed into a product of local generators in a neighborhood that is fully contained within the bulk. Similarly for a decoupled stack of 1D SPT chains, a line symmetry with endpoint operators should be a product of local stabilizers contained within the same 1D chain, and hence not a nonlocal stabilizer. 
\newline

\begin{figure}[t]
\center
	\includeTikz{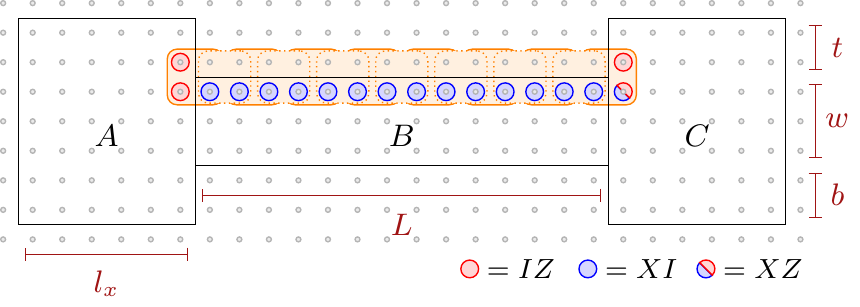}{
		\begin{tikzpicture}[scale=.1]
\def\dy{3}
\def\w{.25}
\def\nx{27}
\def\ny{8}
\def\e{2}
\def\s{.5}
\def\sy{.25}
\def\de{-.25}
	\begin{scope}
			\foreach \x in {7,9,11,...,21} {
			\node[transform shape,draw=orange,fill=orange!80!white!15,regular polygon,regular polygon sides=4,minimum size=8cm,rounded corners=.1cm] at (\x*\dy-.5*\dy,5.5*\dy) {};
			}
			\foreach \x in {8,10,12,...,20} {
			\node[transform shape,draw=orange,densely dotted,fill=orange!80!white!15,regular polygon,regular polygon sides=4,minimum size=7.5cm,rounded corners=.1cm,] at (\x*\dy-.5*\dy,5.5*\dy) {};
			}
		\foreach \x in {7,8,...,20} {
			\node[transform shape,draw=blue,fill=blue!80!white!20,circle,minimum size=1.8cm] at (\x*\dy,5*\dy) {};
        }	
			\node[transform shape,draw=red,fill=red!80!white!20,circle,minimum size=1.8cm] at (6*\dy,5*\dy) {};
			\node[transform shape,draw=red,fill=red!80!white!20,circle,minimum size=1.8cm] at (6*\dy,6*\dy) {};
			\node[transform shape,draw=blue,semicircle,minimum size=.9cm,anchor=south,rotate=135,fill={blue!80!white!20}] at (21*\dy,5*\dy) {};
			\node[transform shape,draw=red,semicircle,minimum size=.9cm,anchor=south,rotate=-45, fill={red!80!white!20}] at (21*\dy,5*\dy) {};
			\node[transform shape,draw=red,fill=red!80!white!20,circle,minimum size=1.8cm] at (21*\dy,6*\dy) {};
			\foreach \x in {0,1,...,\nx} {
        \foreach \y in {0,1,...,\ny} {
		  \draw[color=black!29,fill=black!10!white!80] (\x*\dy,\y*\dy) circle (\w);
        }
        }
	\draw[draw=black,line width=.2pt] (.5*\dy,.5*\dy) -- (\ny*\dy-1.5*\dy,.5*\dy) --  (\ny*\dy-1.5*\dy,\ny*\dy-.5*\dy) -- (.5*\dy,\ny*\dy-.5*\dy) -- cycle 
	(\nx*\dy-.5*\dy,.5*\dy) -- (\nx*\dy-\ny*\dy+1.5*\dy,.5*\dy) --  (\nx*\dy-\ny*\dy+1.5*\dy,\ny*\dy-.5*\dy) -- (\nx*\dy-.5*\dy,\ny*\dy-.5*\dy) -- cycle 
	(-1.5*\dy+\ny*\dy,.5*\dy+\e*\dy) --  (\nx*\dy-\ny*\dy+1.5*\dy,.5*\dy+\e*\dy) 
	(-1.5*\dy+\ny*\dy,\ny*\dy-.5*\dy-\e*\dy) --  (\nx*\dy-\ny*\dy+1.5*\dy,\ny*\dy-.5*\dy-\e*\dy)  ; 
  	\draw[draw=darkred,line width=.2pt,|-|]    (\ny*\dy+\de*\dy-1*\dy,1.5*\dy) to (\nx*\dy -\de*\dy-\ny*\dy+1*\dy, 1.5*\dy);
   	\draw[draw=darkred,line width=.2pt,|-|]    (\ny*\dy-2*\dy-\de*\dy,-.5*\dy) to (1*\dy +\de*\dy, -.5*\dy);
	\node at (3.5*\dy,3.5*\dy){$A$};
	\node at (.5*\nx*\dy,3.5*\dy){$B$};
	\node at (\nx*\dy-3.5*\dy,3.5*\dy){$C$};
	\node at (.5*\nx*\dy,.5*\dy){\red{ $L$}};
	\node at (3.5*\dy,-1.5*\dy){\red{$l_x$}};
   	\draw[draw=darkred,line width=.2pt,|-|]    (\nx*\dy +1*\dy-\s*\dy,.5*\dy+\sy*\dy) to (\nx*\dy +1*\dy-\s*\dy, 2.5*\dy-\sy*\dy);
   	\draw[draw=darkred,line width=.2pt,|-|]    (\nx*\dy +1*\dy-\s*\dy,2.5*\dy+\sy*\dy) to (\nx*\dy+1*\dy-\s*\dy, 5.5*\dy-\sy*\dy);
   	\draw[draw=darkred,line width=.2pt,|-|]    (\nx*\dy +1*\dy-\s*\dy,5.5*\dy+\sy*\dy) to (\nx*\dy +1*\dy-\s*\dy, 7.5*\dy-\sy*\dy);
	\node at (\nx*\dy+1*\dy +.25*\dy,4*\dy){\red{$w$}};
	\node at   (\nx*\dy +1*\dy+.25*\dy,6.25*\dy+\sy*\dy){\red{$t$}};
	\node at (\nx*\dy +1*\dy+.25*\dy,1.25*\dy+\sy*\dy) {\red{$b$}};
	\node[transform shape,draw=red,fill=red!80!white!20,circle,minimum size=1.8cm] at (\nx*\dy-1.4*\ny*\dy,-1*\dy) {};
	\node[anchor=west] at (\nx*\dy-1.4*\ny*\dy+.2*\dy,-1*\dy) {\footnotesize $=IZ$};
	\node[transform shape,draw=blue,fill=blue!80!white!20,circle,minimum size=1.8cm] at (\nx*\dy-1.4*\ny*\dy+4.*\dy,-1*\dy) {};
	\node[anchor=west] at (\nx*\dy-1.4*\ny*\dy+4*\dy+.2*\dy,-1*\dy) {\footnotesize $=XI$};
			\node[transform shape,draw=blue,semicircle,minimum size=.9cm,anchor=south,rotate=135,fill={blue!80!white!20}] at  (\nx*\dy-1.4*\ny*\dy+8*\dy,-1*\dy) {};
			\node[transform shape,draw=red,semicircle,minimum size=.9cm,anchor=south,rotate=-45, fill={red!80!white!20}] at (\nx*\dy-1.4*\ny*\dy+8*\dy,-1*\dy) {};
	\node[anchor=west] at (\nx*\dy-1.4*\ny*\dy+8*\dy+.2*\dy,-1*\dy) {\footnotesize $=XZ$};
	\end{scope}
	\end{tikzpicture}
	}
 \caption{ The regions used to calculate $S_{\text{dumb}}$. A nonlocal stabilizer for Hamiltonian~\eqref{eq:cstate} is depicted (red, blue) along with the support of its local generators (orange).  }
\label{Sdumbbell}
\end{figure}
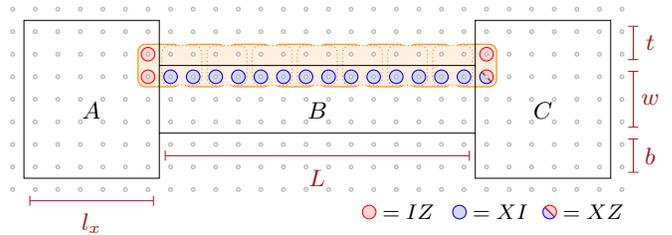

\noindent\textbf{Counting spurious nonlocal stabilizers.}
\label{sec:sdumb}
We now define an entropic quantity $S_{\text{dumb}}$ that is insensitive to 1-form symmetries and decoupled stacks of 1D SPT chains, but is sensitive to rigid string symmetries that lead to spurious contributions to TEE. It is given by the tripartite information of three subsystems whose union forms a dumbbell shape, such as in Fig.~\ref{Sdumbbell}
 \begin{align}
 S_{\text{dumb}} &= I_3(A,B,C) 
 \\
 & \approx - I(A:C|B)
  = S_B + S_{ABC} - S_{AB} - S_{BC}
  \nonumber
 \, ,
 \end{align}
which is approximately equal to the mutual information between $A$ and $C$, conditioned on $B$, since ${S_{A}+S_{C}-S_{AC}\approx 0}$ when $L$ is much larger than the correlation length.  
Hence for short range correlated states $S_{\text{dumb}} \leq 0$, by the strong subbaditivity of quantum entropy~\cite{doi:10.1063/1.1666274}, and the stability of TEE result in Ref.~\onlinecite{PhysRevB.86.245116} (see also Ref.~\onlinecite{PhysRevB.97.144106}) is sidestepped when $S_{\text{dumb}} < 0 $. 

If one assumes the entropy of a region takes the form given in Eq.~\eqref{eq:arealaw}, plus possible corner contributions due to the lattice cutoff $\epsilon$, one finds $S_{\text{dumb}}\approx 0$. A nonzero value arises due to nonlocal constraints on the $ABC$ subsystem that reduce $S_{ABC}$ but are not captured by a constant correction to the area law. 

For general gapped local Hamiltonians with a nonzero (but finite) correlation length $\xi$, $S_\text{dumb}$ counts the number of independent nonlocal constraints on a dumbbell shaped region.
Let us assume that all lengths $L$,  $l_x$, $t$, $b$ are large compared to $\xi$. For simplicity further assume the dumbbell is oriented along the direction of a 1D rigid subsystem symmetry $G=\langle \Gamma \rangle$, where $\Gamma$ is a set of independent generators, and the endpoint annihilation operators act nontrivially on a disc of radius $\xi$. Then we can estimate $S_\text{dumb} \approx - 2 |\Gamma| \xi/\epsilon$, since for each generating group element, there will be $2 \xi/\epsilon$ independent nonlocal constraints that cannot be deformed to lie within $AB, \, BC,$ or $AC,$ corresponding to segments of rigid string symmetry with endpoint operators that are ``trapped'' within $A$ and $C$, see Fig.~\ref{Sdumbbell} for example. If the support of the endpoint operators is smaller than $\xi$ the value of $S_\text{dumb}$ will be closer to $0$. 
For the cluster state example we find $S_\text{dumb}=-2$, for $t,b>0$, since each generator in $\Gamma$ only contributes a single nonlocal stabilizer due to the particular shape of the endpoint operators. 
Even in the more general setting, 1-form SPTs and stacked decoupled 1D SPTs should not contribute to  $S_\text{dumb}$ as all constraints in the region $ABC$ can be deformed into $A$ or $C$ for such models.

Although $S_\text{dumb}$ is not fully deformation invariant, as it depends strongly on the directions of the subsystem symmetries, it should not depend on the precise length of $L$,  $l_x$, $t$, and $b$, provided they are large compared to the correlation length. 

The value of $S_\text{dumb}$ may jump under symmetry-respecting deformations of the Hamiltonian or the application of symmetry-respecting local unitary circuits. For a generic such deformation, which is assumed to increase the correlation length $\xi$, the value of $S_\text{dumb}$ should monotonically decrease. 
For small nonsymmetric Hamiltonian deformations we expect $S_\text{dumb}$ to decay exponentially with $L$, controlled by the \emph{string correlation length} (see below), but remain insensitive to changes in $l_x$, $t$, and $b$, see the appendix.

We now argue that the spurious contributions to $S_\text{topo}$ are captured by the value of $S_\text{dumb}$.  For $S^{\text{LW}}_{\text{topo}}$, which is approximately given by a conditional mutual information in a similar fashion to $S_\text{dumb}$, the excess spurious nonlocal stabilizers in $ABC$ correspond to the same rigid stringlike symmetries with ``trapped'' endpoints that contribute to $S_\text{dumb}$, see Fig.~\ref{clusterlattice}~c) for example. 
For $S^{\text{KP}}_{\text{topo}}$, one can understand the excess spurious tripartite information as being due to a certain kind of tripartite entanglement that also contributes to $S_\text{dumb}$.
Consequently $S_\text{topo}$ may not be local unitary invariant or deformation invariant when $S_\text{dumb}\neq 0$. 
For the cluster state example $S_\text{dumb}=-1$, for $t=0$, captures the nonlocal stabilizers that lead to $\Delta S_B - \Delta S_{AB} =1$ under the deformations shown in Fig.~\ref{clusterlattice}. More generally $S_\text{dumb}$, for $t,b>0$, captures nonlocal stabilizers that cause $S_{\text{topo}}$ to jump under deformations of a region's boundary that runs along the direction of a subsystem symmetry. Hence $S_\text{dumb} < 0$ for a stabilizer model implies the conditions needed to derive $\gamma$ from $S_\text{topo}$ are not met. 
\newline

\noindent\textbf{String correlation length and SSPT.}
A consequence of a nonzero value of $S_\text{dumb}$ is the existence of an infinite string correlation length. 
The string correlation length $ \xi_\text{string}$ for a symmetry operator $g_{1,L}$ of length $L$ is defined by 
\begin{align}
\label{eq:strcorlen}
\max_{O_1,  O_L } \left(  \langle  O_1\, g_{1,L}\,  O_L \rangle - \langle O_1 \rangle \langle O_L \rangle \right) \sim e^{- { L / \xi_\text{string}}}
\end{align}
where $O_1$ and $O_L$ are operators within a neighborhood of the string's end points. For our cluster state example the $g_{1,L}$ are given by restricting the symmetry generators in Eq.~\eqref{eq:subgen} to $L$ sites, and  $O_1,O_L$ can be chosen from Eq.~\eqref{eq:csendpointops} to attain $ \xi_\text{string}=\infty$. 

The presence of an infinite rigid string correlation length is a 
defining property of an SSPT phase, since a 2D SSPT on a long cylinder can be viewed as a 1D (on-site) SPT\footnote{
This mapping respects the symmetries but not the locality in the compactified direction. An infinite string correlation function for the 1D SPT implies one for the 2D SSPT since 
the maximum in Eq.~\eqref{eq:strcorlen} is attained by end point operators respecting the 2D locality.
} 
and such phases are characterized by infinite string correlation functions~\cite{PhysRevB.45.304,pollmann2010entanglement,Else2012b}. 
Furthermore, it should persist under symmetric perturbations until a phase transition is crossed. However, under an arbitrary nonsymmetric perturbation this property is reduced to a finite string correlation length. 

Conversely, any SSPT whose edge projective representation~\cite{you2018subsystem,subsystemphaserel} cannot be trivialized by changing the projective representations of symmetry groups on individual rows leads to a nonzero $S_\text{dumb}$, which remains nonzero under symmetry-respecting local unitary transformations. This is because the end point operators of symmetries on any given row must have support on a distinct row to be consistent with such a nontrivial projective representation (as it cannot be abelian). 
So-called strong SSPT phases always satisfy the above condition, see Ref.~\onlinecite{subsystemphaserel} for further details. 
We remark that the condition above is similar to a previously identified~\cite{PhysRevB.94.075151} sufficient condition for spurious TEE via cylinder extrapolation. 
\newline

\noindent\textbf{More examples with spurious TEE.} 
We have examined $S^{\text{KP}}_\text{topo}$ and $S_\text{dumb}$ for a family of states that interpolate between the product state and the cluster state defined in Eq.~\eqref{eq:cstate}, obtained by replacing the $CZ$ gates by controlled-$e^{i \theta}$ gates. While they have zero correlation length, the string correlation length is finite for $\theta< \pi$. For these deformed states, we have found a good agreement between $S^{\text{KP}}_\text{topo}$ and $S_\text{dumb}$ for all values of $\theta$. As the deformation breaks the subsystem symmetries, we expect that $S_\text{dumb}$ (and also $S_\text{topo}$) should decay with $L$, which is indeed observed in the numerics. We have also verified numerically that $S_{\text{dumb}}$ remains almost constant for $w,l_x\rightarrow \infty$ for generic values of $0<\theta<\pi$, see the appendix for more details. 

We have further shown that ${S^{\text{KP}}_\text{topo}-S_\text{dumb}=-\gamma}$ for a family of 2D stabilizer models with spurious TEE, obtained from compactifications of Haah's cubic code~\cite{haah2011local,PhysRevLett.107.150504,bravyi2013quantum,compactifyinghaahcode} on a 2D slab. For details about these, and other, examples see the appendix. 
These results are consistent with the general considerations, and support our proposal that $S_\text{dumb}$ captures the spurious contribution to TEE.
\newline

\noindent\textbf{ Discussion and conclusions.}
\label{sec:conclusion}
We found that bulk calculations of the TEE following Kitaev and Preskill~\cite{KitaevPreskill}, or Levin and Wen~\cite{levinwenentanglement}, may suffer from spurious nontopological contributions. We offered an explanation for such contributions in terms of nonlocal constraints originating from SSPT phases of matter~\cite{you2018subsystem,subsystemphaserel} and introduced an entropic quantity $S_{\text{dumb}}$ to measure them. 
This leads us to conclude that not only the usual correlation length but all (rigid) string correlation lengths should be small, compared to the size of the regions involved in the calculation of $S_\text{topo}$, if the TEE $\gamma$ is to be reliably extracted (unless an alternate approach is used~\cite{Kato2018b,Kato2018a}).  

Our results imply that the functional form of the entanglement entropy for a region $R$, given in Eq.~\eqref{eq:arealaw}, does not hold for states with nontrivial SSPT order. Furthermore, the addition of constant corner terms to Eq.~\eqref{eq:arealaw} does not suffice to explain the spurious values of TEE we have found. To correctly account for the spurious TEE present in nontrivial SSPT phases, which have an infinite string correlation length, one must add a correction term to Eq.~\eqref{eq:arealaw} that depends on the shape of the whole boundary $\partial R$ in a nonlocal way. We remark that if the string correlation length is finite, then these correction terms become quasi-local (with magnitude decaying exponentially with the size of the feature that causes them). 

The spurious contributions to the TEE we have identified are possible in higher dimensions and affect proposals to extract information about fracton topological order from the TEE~\cite{PhysRevB.97.125101}, as discussed in the appendix.
A related phenomenon can also occur in one spatial dimension~\cite{bravyiunpublished,PhysRevB.94.075151}, it would be interesting to see if this can cause spurious contributions to the boundary entropy~\cite{Affleck1991}. 
\newline

\noindent\textbf{Acknowledgements.} 
DW thanks Trithep Devakul, Steve Flammia, and Yizhi You for enlightening discussions. AD thanks Kevin Slagle and Jeongwan Haah for useful comments and discussions. DW and MC are supported by startup funds at Yale University. 
\bibliographystyle{apsrev4-1}
\bibliography{Bibliography}

\onecolumngrid
\clearpage 

\appendix

\section{Stabilizer Hamiltonians}
\label{sec:stabHam}
In this work we have mainly considered stabilizer Hamiltonians. For a stabilizer Hamiltonian 
 \begin{align}
 \sum_{i \text{ lattice sites}} ( \openone - h_i )
 \, ,
 \end{align}
 the terms (generators) $h_i$ must be local Hermitian Pauli operators that act on a neighborhood of site $i$ and mutually commute. 
Pauli operators are tensor products of single qubit Pauli operators: 
\begin{align}
I &=  \begin{pmatrix}
1 & 0 \\
0 & 1
\end{pmatrix}
\, ,
&X &= \begin{pmatrix}
0 & 1 \\
1 & 0
\end{pmatrix}
\, ,
& 
Z &= \begin{pmatrix}
1 & 0 \\
0 & -1
\end{pmatrix}
\, ,
&Y &= \begin{pmatrix}
0 & -i \\
i & 0
\end{pmatrix}
\, .
\end{align}
 The Hamiltonian terms generate the stabilizer subgroup $G:=\langle \{ h_i \}_{i} \rangle $, which should not contain a multiple of the identity $\alpha \openone$ with $\alpha \neq 1$.  
The models considered in this paper are defined on a square lattice with multiple qubits per site and interactions terms acting on sites adjacent to each plaquette. 
The projector onto the ground space is given by 
\begin{align}
\rho := \prod_{i} \frac{1}{2} (\openone + h_i) = \frac{1}{ |G|} \sum_{g\in G} g
\, ,
\end{align}
where we have assumed that there are no \emph{relations} between the local stabilizer generators $h_i$ (products nontrivially equal to the identity) for simplicity. 
By taking the trace of both sides we find $|G|=2^{N}$, where we have further assumed that there is a unique ground state. The result is unchanged if there is a topological degeneracy and an arbitrary ground state is chosen. 

We define the stabilizer group for any subregion of the lattice $R$ to be the subgroup $G_R \leq G$ of stabilizers that act as identity on the exterior of $R$. It is easily verified that the reduced density matrix on region $R$ is given by 
\begin{align}
\rho_R = \tr_{R^c} \, \rho = \frac{1}{ 2^{N_R}} \sum_{g\in G_R} g
\, ,
\end{align}
since the trace of any nontrivial Pauli operator acting on $R^c$ is $0$ while the trace of the identity on $R^c$ is $(N-N_R)$, where $N_R$ is the number of qubits within region $R$.

The entanglement R\'enyi entropies are defined as follows 
\begin{align}
\label{eq:renyientropy}
S^{(\alpha)}(\rho)=\frac{1}{1-\alpha} \log \tr \rho^\alpha
\, ,
\end{align}
for $\alpha \rightarrow 1$ we recover the von Neumann entropy. 
The entanglement entropy of a stabilizer state across the boundary of a region $R$ is given by~\cite{fattal2004entanglement,PhysRevA.71.022315}  
\begin{align}
S^{(\alpha)}(\rho_R)= N_R - \log_2 |G_R|
\, ,
\end{align}
for all $\alpha$, since the spectrum of $\rho_R$ is flat. 
When calculating the entanglement entropy of a stabilizer model it is important to ensure that all relations have been accounted for, and that all \emph{nonlocal stabilizers} in $G_R$ (those that are supported on $R$ but are not products of the local generators contained in $R$) are included in the counting~\cite{PhysRevB.94.075151,PhysRevB.97.125101}. 
In this case the choice of  R\'enyi entropy clearly does not matter. More generally it has been found that for topological properties, at long distances, the choice of R\'enyi entropy is unimportant~\cite{topologicalrenyi,PhysRevA.86.062330}. For the sake of ease, we have focused on $\alpha=2$ in our calculations for nonstabilizer models, where such a choice must be made.

A CSS code~\cite{PhysRevA.54.1098,Steane2551} Hamiltonian is a stabilizer Hamiltonian where the generators $h_i$ are all tensor products of either Pauli $X$ or $Z$ operators. It has been shown that \emph{translation invariant local CSS code Hamiltonians in 2D, satisfying a topological order condition, can be disentangled into a tensor product of toric code Hamiltonians, along with trivial Hamiltonians, via some choice of local Clifford circuit}~\cite{bombin2014structure,haah2016algebraic}. Hence the TEE is a complete invariant for the topological phase of such models, as it counts the number of copies of toric code. 
An implication one can draw from the current work is that in the presence of subsystem symmetry this result must be refined. In particular if one demands that the local unitaries respect the subsystem symmetry one may find copies of toric code, possibly enriched by the subsystem symmetry, along with potentially nontrivial SSPT states.

\section{Further examples}
\label{sec:furtherexamples}

\subsection{Stacked 1D cluster states}
In this section we introduce a model consisting of stacked 1D cluster states. 
We find that the model does not give rise to nonzero  values of  $S_\text{topo}$ or $S_\text{dumb}$, however, this changes after a symmetric local unitary circuit is applied.

The local stabilizer generators are given by 
 \begin{align}
 \label{eq:stackedcluster}
 \begin{array}{c}
 \plaquette{II}{II}{IZ}{XZ}
 \quad
 \plaquette{II}{II}{ZX}{ZI}
 \end{array}
 \, ,
\end{align}
this model is a weak SSPT. 
We find $S^{\text{KP}}_\text{topo}=S^{\text{LW}}_\text{topo}=0$ and no correction via the cylinder extrapolation method. 
We also find $S_\text{dumb}=0$. However, as discussed above, these quantities are not invariant under symmetry respecting local unitary circuits. For example, we may apply a product of $(H\otimes H) CZ  (H\otimes H) $ gates along the vertical direction of the lattice and in the resulting model we find $S_\text{dumb}=-4$, ${S^{\text{KP}}_\text{topo}=S^{\text{LW}}_\text{topo}=-4}$, and a $-2$ correction from cylinder extrapolation. 
For completeness we define the gates used above: 
\begin{align}
H = \begin{pmatrix}
1 & 1 \\
1 & -1
\end{pmatrix}\, , \qquad CZ =  
 \begin{pmatrix}
 1 & 0 & 0 & 0 \\
0  & 1 & 0 & 0  \\
0  & 0 & 1 & 0 \\
0  & 0 & 0 & -1
 \end{pmatrix}
\, .
\end{align}

\subsection{Zigzagging 1D cluster states}
 
Here, we consider a model consisting of 1D cluster states stacked in a nontrivial zigzagging pattern. Although this model is classified as a weak SSPT, according to the definition in Ref.~\onlinecite{subsystemphaserel}, we show that it gives rise to nontrivial $S_\text{topo}$ and $S_\text{dumb}$ quantities. We make sense of this result by demonstrating that it is related to the 2D cluster state via a decoupled stack of 1D local unitary circuits, which do not change the result of our $S^{\text{LW}}_\text{topo}$ or the horizontal $S_\text{dumb}$. 
We remark that such zigzagging SPT states were previously noted to cause spurious TEE on the cylinder~\cite{PhysRevB.92.075104,PhysRevB.94.075151}. 

The local stabilizer generators for a stack of zigzagging cluster states are given by translates of 
 \begin{align}
 \label{eq:zigzagcluster}
 \begin{array}{c}
 \plaquette{IZ}{IZ}{II}{XI}
 \quad
 \plaquette{IX}{II}{ZI}{ZI}
 \end{array}
 \, .
\end{align}
For this model we find $S^{\text{KP}}_\text{topo}=-1$, $S^{\text{LW}}_\text{topo}=-2$, $S_\text{dumb}=-2$,  and a $-1$ correction from the cylinder extrapolation method. We remark that one would find no spurious TEE from both Levin-Wen and cylinder extrapolation if the system is rotated by 90 degrees.  

We make sense of these results by noting that the zigzagged cluster state is mapped to the 2D cluster state via a decoupled stack of 1D local unitary circuits, built from a product of $CZ$ gates applied to nearest neighbor qubits along a row. 
One can verify that such a circuit does not effect the result of  $S^{\text{LW}}_\text{topo}$ or $S_\text{dumb}$, aligned along the horizontal axis. Although the circuit is not locally subsystem symmetric, each decoupled 1D row is globally symmetric under the horizontal subsystem symmetries. We anticipate that such operations should be allowed in the relatively loose equivalence relations for SSPT phases~\cite{subsystemphaserel}. Hence we expect the zigzagged cluster state and the 2D cluster state to lie in the same SSPT phase with respect to the horizontal subsystem symmetries only.

\subsection{Global and 1-form symmetric cluster state}

We consider a cluster model with a global $\Z_2$, and a 1-form $\Z_2$ symmetry, with local stabilizer generators 
 \begin{align}
 \label{eq:1fcluster}
 \begin{array}{c}
 \plaquettedoublespacing{IIZ}{ZXZ}{}{ZII}
 \quad
 \plaquettedoublespacing{}{}{IZX}{IZI}
   \quad
 \plaquette{IZI}{}{XZI}{}
 \end{array}
 \, ,
\end{align}
where the first qubit on each vertex corresponds to the positive adjacent vertical edge, and the third to the positive adjacent horizontal edge. 
For this model we find $S^{\text{KP}}_\text{topo}=S^{\text{LW}}_\text{topo}=0$,  $S_\text{dumb}=0$, and no correction via cylinder extrapolation. 

\subsection{Compactified cubic code}

In this section we consider a family of models for which ${-S^{\text{KP}}_\text{topo}-\gamma}$ diverges. We remark these models are not simply constructed by stacking other known examples. 

The simplest such model has $6$ qubits per site of a square lattice that is specified by translates of the local stabilizer generators
 \begin{align}
 \label{eq:lz3model}
 \begin{array}{c}
  \plaquettedoublespacing{X_1X_2X_4}{X_2X_3}{X_2X_3}{X_1}
 \quad
   \plaquettedoublespacing{X_3X_4X_6}{X_4X_5}{X_4X_5}{X_3}
 \quad 
   \plaquettedoublespacing{X_2X_5X_6}{X_1X_6}{X_1X_6}{X_5}
 \\
  \plaquettedoublespacing{Z_4}{Z_2Z_3}{Z_2Z_3}{Z_1Z_3Z_4}
   \quad
  \plaquettedoublespacing{Z_6}{Z_4Z_5}{Z_4Z_5}{Z_3Z_5Z_6}
  \quad
 \plaquettedoublespacing{Z_2}{Z_1Z_6}{Z_1Z_6}{Z_1Z_2Z_5}
 \end{array}
 \, .
\end{align}
This model is in the same topological phase as two copies of toric code, the fundamental excitations correspond to violations of single plaquette terms. The string operators are given by $X_1 X_2 X_3 X_6,\, Z_1 Z_2 Z_3 Z_6$ and $X_2 X_3 X_4 X_5, \, Z_2 Z_3 Z_4 Z_5$ which create pairs of excitations on a single plaquette and 
\begin{align}
   \plaquettedoublespacing{X_2 X_3 X_4 X_5}{}{}{X_3}
   \quad 
      \plaquettedoublespacing{X_1 X_2 X_3 X_6}{}{}{X_4 X_5}
      \quad
   \plaquettedoublespacing{Z_2}{}{}{Z_1 Z_4 Z_5 Z_6}
   \quad 
      \plaquettedoublespacing{Z_1 Z_6}{}{}{Z_2 Z_3 Z_4 Z_5}
\, ,
\end{align}
which create pairs of excitations on two corner sharing plaquettes. If the plaquettes are labeled with a checkerboard, different superselection sectors live on different colours. Translation acts to permute the anyons in a nontrivial manner, and one can verify that this model exhibits symmetry-enriched topological order corresponding to two copies of toric code with a nontrivial $\mathbb{Z}_2$ action under translation, see Ref.~\onlinecite{compactifyinghaahcode} for further details.  

We find $S^{\text{KP}}_{\text{topo}}=-6$, as there are no local relations amongst the stabilizer generators and no nonlocal stabilizers for rectangular regions. We also find $S^{\text{LW}}_{\text{topo}}=-8$. The cylinder extrapolation must be carried out with care in this case, since odd cylinder circumferences correspond to twisted boundary conditions, we consider a circumference $L=2 n,\, n\in \mathbb{Z}^+$ and pick out the vacuum minimally entangled state by including the anyonic string operators that wind the cylinder. For $n= 3 k, \, k\in \mathbb{Z}^+$ we find  a correction of $-4$, while for other $n$ we find  a correction of $-2$. 

The model defined in Eq.~\eqref{eq:lz3model} has a pair of $\Z_2 \times \Z_2$ symmetries along each line in the $x$ and $y$ direction, one consisting of $X$ operators and another consisting of $Z$ operators. 
\begin{align}
&\bigotimes_i (X_1 X_2 X_3 X_4 )_{3i,j} ( X_3 X_4 X_5 X_6 )_{3i+1,j} (X_1 X_2 X_5 X_6 )_{3i+2,j}
\, ,
&\bigotimes_i& (X_1 X_2 X_3 X_4 )_{i,3j} ( X_3 X_4 X_5 X_6 )_{i,3j+1} (X_1 X_2 X_5 X_6 )_{i,3j+2}
\, ,
\nonumber
\\
&\bigotimes_i (Z_1 Z_2 Z_3 Z_4 )_{3i,j} ( Z_3 Z_4 Z_5 Z_6 )_{3i+1,j} (Z_1 Z_2 Z_5 Z_6 )_{3i+2,j}
\, ,
&\bigotimes_i& (Z_1 Z_2 Z_3 Z_4 )_{i,3j} ( Z_3 Z_4 Z_5 Z_6 )_{i,3j+1} (Z_1 Z_2 Z_5 Z_6 )_{i,3j+2}
\, .
\end{align}
We remark that translations along the orientation of these line symmetry act nontrivially on them, and hence the total group acting on each line is a nontrivial extension of the subsystem symmetries by translation. 

For the model in Eq.~\eqref{eq:lz3model} the endpoint operators depend on the string symmetry operator considered, there are three distinct options for $X$ and $Z$ horizontal strings, respectively, given by
 \begin{align}
 \begin{array}{c}
 \xymatrix@!0{%
X_1 X_2 X_3 X_6 \ar@{-}[d]  
\\
X_1 X_5 
}
\end{array}
\, ,
&&
  \begin{array}{c}
 \xymatrix@!0{%
X_2 X_3 X_4 X_5  \ar@{-}[d]  
\\
X_1 X_3 
}
 \end{array}
 \, ,
 &&
  \begin{array}{c}
 \xymatrix@!0{%
X_1 X_4 X_5 X_6   \ar@{-}[d]  
\\
X_3 X_5
}
 \end{array}
 \, ,
 &&
  \begin{array}{c}
 \xymatrix@!0{%
Z_2 Z_4 \ar@{-}[d]  
\\
Z_1 Z_2 Z_3 Z_6
}
\end{array}
\, ,
&&
  \begin{array}{c}
 \xymatrix@!0{%
Z_4 Z_6  \ar@{-}[d]  
\\
Z_2 Z_3 Z_4 Z_5
}
 \end{array}
 \, ,
 &&
  \begin{array}{c}
 \xymatrix@!0{%
Z_2 Z_6 \ar@{-}[d]  
\\
Z_1 Z_4 Z_5 Z_6
}
 \end{array}
 \, ,
\end{align}
and similar operators, rotated by $90$ degrees, for vertical strings. 

The above example is actually the smallest instance of a family of examples that arise from compactifications of Haah's cubic code~\cite{haah2011local,PhysRevLett.107.150504,bravyi2013quantum,compactifyinghaahcode}. 
These models live on a square lattice with $2q$ qubits per site, for $q\in 3 \Z^+$. The stabilizer generators are given by translations of the following $2q$ plaquette terms 
 \begin{align}
 \begin{array}{c}
 \qquad\plaquettetriplespacing{Z_{4+2i}}{Z_{2+2i}Z_{3+2i}}{Z_{2+2i}Z_{3+2i}}{Z_{1+2i}Z_{3+2i}Z_{4+2i}}
 \qquad
  \plaquettetriplespacing{X_{2i+1}X_{2i+2}X_{2i+4}}{X_{2i+2}X_{2i+3}}{X_{2i+2}X_{2i+3}}{X_{2i+1}}
 \end{array}
 \, ,
\end{align}
where $i\in \Z_q$ (additions are mod $2q$). 
For this family of examples we find $S^{\text{KP}}_{\text{topo}}=-2q$, while the respective model lies in the same phase as  $<2q$ copies of toric code~\cite{compactifyinghaahcode}. 
Specifically, for $q=3\cdot 2^{i}$. Using the results of the bifurcating real-space renormalization group procedure in Ref.~\onlinecite{haah2014bifurcation}, the number of toric codes is $2^{i+1}$, leading  to a discrepancy of $2^{i+2}$, see Ref.~\onlinecite{compactifyinghaahcode} for further details.

For the models in this family, with $q\leq21$, we have numerically verified that $S^{\text{KP}}_\text{topo}-S_\text{dumb}=-\gamma$.

\subsection{3D linear subsystem symmetric cluster state}

In this section we demonstrate by example that spurious contributions to the TEE can occur in three and higher dimensions. We first introduce regions used to calculate TEE in 3D followed by a 3D cluster model that leads to spurious TEE. 

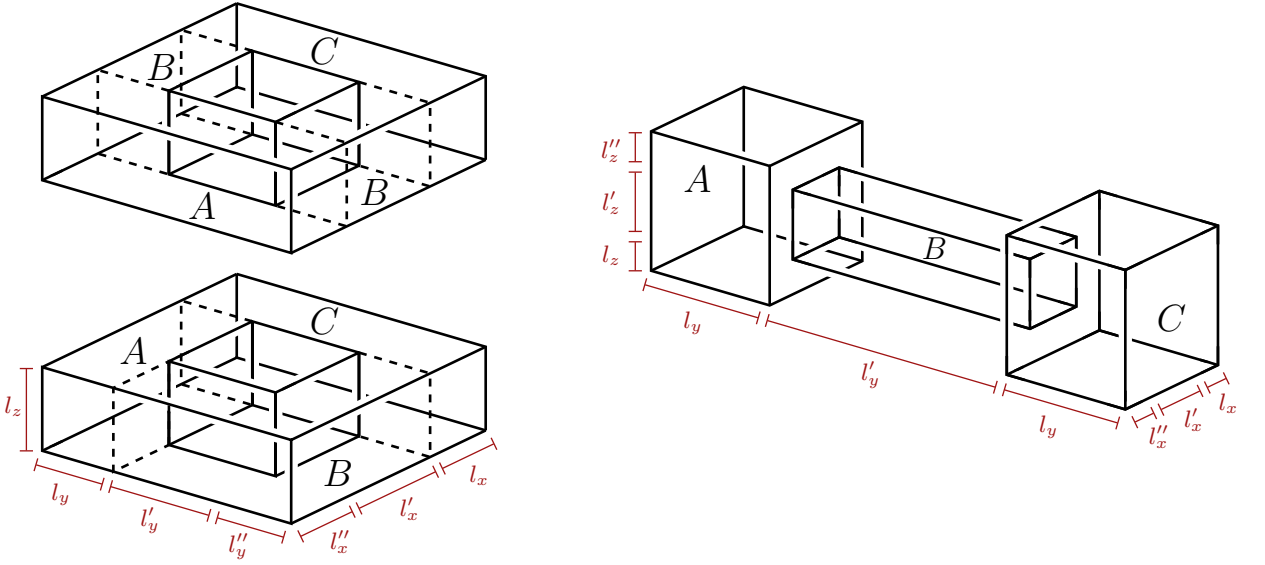
\begin{figure}[t]
\center
\begin{subfigure}{0.49\linewidth}
	\includeTikz{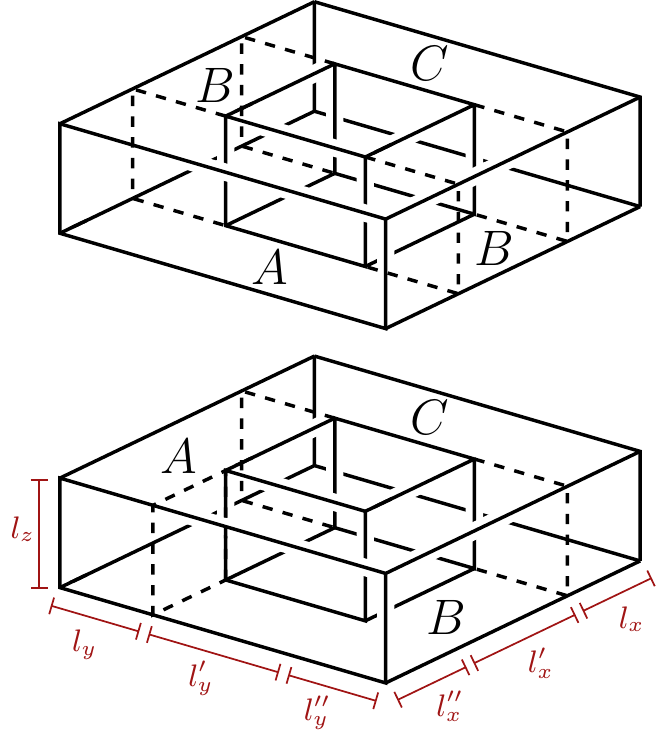}{
	\tdplotsetmaincoords{68}{128}
	\begin{tikzpicture}[scale=.6,tdplot_main_coords]
	\def\dx{.25}
	\def\de{.1}
	\def\dl{.6}
	
 	 \draw[ black, line width = 1pt] 
 	  	  (0,0,2) -- (0,0,0) -- (0,7,0)  
  	  	 (0,0,0) -- (7,0,0);

 	 \draw[ white, double = black, line width = 2pt, double distance = 1pt ] 
 	  	 (2,2,0) -- (2,5,0) -- (2,5,2) -- (2,2,2) -- cycle
		(2,2,0) -- (2,2,2) 
		(2,5,0) -- (2,5,2) 
		(5,2,2) -- (2,2,2) 
	  	 (5,5,2) -- (2,5,2);

 	 \draw[ white, double = black, line width = 2pt, double distance = 1pt ] 
 	  	 (5,2,0) -- (5,5,0) -- (5,5,2) -- (5,2,2) -- cycle 
 	  	 (5,2,0) -- (5,2,2) 
 	  	 (5,5,0) -- (5,5,2);

	\draw[ black, line width = 1pt] 
		(2,2,0) -- (5,2,0)  
		(2,5,0) -- (5,5,0)  ;

 	 \draw[ white, double = black, line width = 2pt, double distance = 1pt ] 
 	  	 (7,0,0) -- (7,7,0) -- (7,7,2) -- (7,0,2) -- cycle
	  	 (0,7,2) -- (7,7,2);

	\draw[ black, line width = 1pt] 
 	 (0,7,0) -- (7,7,0)
 	 (0,0,2) -- (7,0,2)
	(0,7,0) -- (0,7,2) -- (0,0,2);

	\begin{scope}	
	\clip (5,2,2) circle (.5cm)
	(5,5,2) circle (.5cm)
	(7,0,0) circle (.5cm);	
	\draw[ black, line width = 1pt] 
		(5,2,2) -- (2,2,2)  ;	
	\draw[ black, line width = 1pt] 
		(5,5,2) -- (2,5,2)  ;	
	\draw[ black, line width = 1pt] 
		(7,0,0) -- (0,0,0)  ;	
	\end{scope}
	
	\draw[ black, dashed, line width = 1pt] 
 	 (2,2,2) -- (2,0,2) -- (2,0,0) -- (2,2,0) 
 	 (5,2,2) -- (5,0,2) -- (5,0,0) -- (5,2,0) 
 	 (2,5,2) -- (2,7,2) -- (2,7,0) -- (2,5,0) 
 	 (5,5,2) -- (5,7,2) -- (5,7,0) -- (5,5,0);
	
	\node at (7,4.5,.5) {\Large $A$};
	\node at (4,1,2) {\Large $B$};
	\node at (4,7,.5) {\Large $B$};
	\node at (1,3.25,2) {\Large $C$};
	
%

	\begin{scope}[yshift=-6cm] 
	
	 	 \draw[ black, line width = 1pt] 
 	  	  (0,0,2) -- (0,0,0) -- (0,7,0)  
  	  	 (0,0,0) -- (7,0,0);

 	 \draw[ white, double = black, line width = 2pt, double distance = 1pt ] 
 	  	 (2,2,0) -- (2,5,0) -- (2,5,2) -- (2,2,2) -- cycle
		(2,2,0) -- (2,2,2) 
		(2,5,0) -- (2,5,2) 
		(5,2,2) -- (2,2,2) 
	  	 (5,5,2) -- (2,5,2);

 	 \draw[ white, double = black, line width = 2pt, double distance = 1pt ] 
 	  	 (5,2,0) -- (5,5,0) -- (5,5,2) -- (5,2,2) -- cycle 
 	  	 (5,2,0) -- (5,2,2) 
 	  	 (5,5,0) -- (5,5,2);

	\draw[ black, line width = 1pt] 
		(2,2,0) -- (5,2,0)  
		(2,5,0) -- (5,5,0)  ;

 	 \draw[ white, double = black, line width = 2pt, double distance = 1pt ] 
 	  	 (7,0,0) -- (7,7,0) -- (7,7,2) -- (7,0,2) -- cycle
	  	 (0,7,2) -- (7,7,2);

	\draw[ black, line width = 1pt] 
 	 (0,7,0) -- (7,7,0)
 	 (0,0,2) -- (7,0,2)
	(0,7,0) -- (0,7,2) -- (0,0,2);
	
	\begin{scope}
	\clip (5,2,2) circle (.5cm)
	(5,5,2) circle (.5cm)
	(7,0,0) circle (.5cm);	
	\draw[ black, line width = 1pt] 
		(5,2,2) -- (2,2,2)  ;	
	\draw[ black, line width = 1pt] 
		(5,5,2) -- (2,5,2)  ;	
	\draw[ black, line width = 1pt] 
		(7,0,0) -- (0,0,0)  ;	
	\end{scope}
	
	\draw[ black, dashed, line width = 1pt] 
 	 (2,2,2) -- (2,0,2) -- (2,0,0) -- (2,2,0)
 	 (2,5,2) -- (2,7,2) -- (2,7,0) -- (2,5,0)
  	(5,2,0) -- (5,2,2) -- (7,2,2) -- (7,2,0) -- cycle;

	\node at (5,1,2) {\Large $A$};
	\node at (5,6.75,.5) {\Large $B$};
	\node at (1,3.25,2) {\Large $C$};

  	\draw[draw=darkred,line width=.5pt,|-|]    (7+\dx,0-\dx,0) to  (7+\dx,0-\dx,2);
   	\draw[draw=darkred,line width=.5pt,|-|]    (7+\dx,0,0-\dx) to  (7+\dx,2-\de,0-\dx);
   	\draw[draw=darkred,line width=.5pt,|-|]    (7+\dx,2+\de,0-\dx) to  (7+\dx,5-\de,0-\dx);
   	\draw[draw=darkred,line width=.5pt,|-|]    (7+\dx,5+\de,0-\dx) to  (7+\dx,7,0-\dx);
   	\draw[draw=darkred,line width=.5pt,|-|]    (7,7+\dx,0-\dx) to  (5+\de,7+\dx,0-\dx);
   	\draw[draw=darkred,line width=.5pt,|-|]    (5-\de,7+\dx,0-\dx) to  (2+\de,7+\dx,0-\dx);
   	\draw[draw=darkred,line width=.5pt,|-|]    (2-\de,7+\dx,0-\dx) to  (0,7+\dx,0-\dx);
   	
	\node at (1,7+\dl,0-\dl){\red{$l_x$}};
	\node at (3.5,7+\dl,0-\dl){\red{$l_x'$}};
	\node at (6,7+\dl,0-\dl){\red{$l_x''$}};
	\node at (7+\dl,6,0-\dl){\red{$l_y''$}};
	\node at (7+\dl,3.5,0-\dl){\red{$l_y'$}};
	\node at (7+\dl,1,0-\dl){\red{$l_y$}};
	\node at (7+\dx,0-\dl,1){\red{$l_z$}};
	\end{scope}
	
	\end{tikzpicture}
	}
\end{subfigure}
\begin{subfigure}{0.49\linewidth}
	\includeTikz{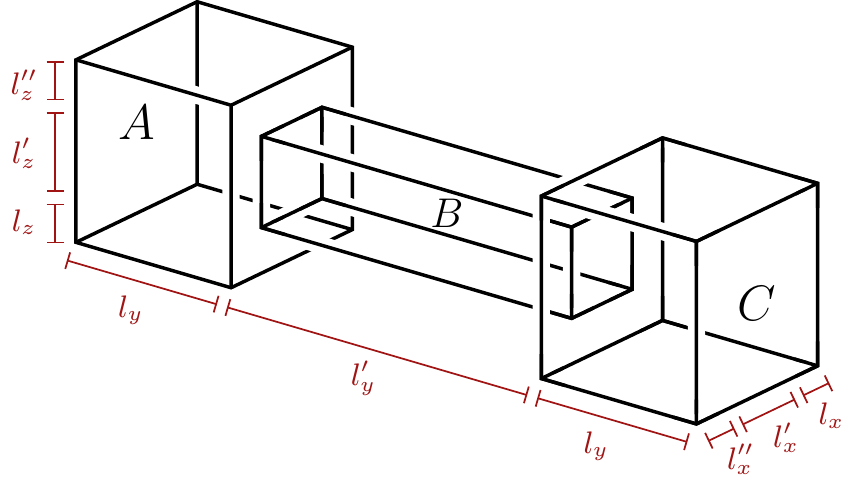}{
	\tdplotsetmaincoords{68}{128}
	\begin{tikzpicture}[scale=.5,tdplot_main_coords]

	\def\l{8}
	\def\dx{.3}
	\def\de{.13}
	\def\dl{.75}
	
	\draw[ black, line width = 1pt] 
		(0,0,0) -- (0,0,4) 
		(0,0,0) -- (0,4,0)
		(0,0,0) -- (4,0,0);	
	\draw[ white, double = black, line width = 2pt, double distance = 1pt ] 
		(4,4,4) -- (4,4,0)
		(4,4,4) -- (4,0,4)
		(4,4,4) -- (0,4,4);
	\draw[ black, line width = 1pt] 	
		(4,4,0) -- (4,0,0) -- (4,0,4) -- (0,0,4) -- (0,4,4) -- (0,4,0) -- cycle;

	\draw[ white, double = black, line width = 2pt, double distance = 1pt ] 
		(1,4,1) -- (3,4,1)
		(1,4,1) -- (1,4,3)
		(1,4,1) -- (1,4+\l,1);
	\draw[ white, double = black, line width = 2pt, double distance = 1pt ] 
		(3,4+\l,3) -- (3,4,3)
		(3,4+\l,3) -- (1,4+\l,3)
		(3,4+\l,3) -- (3,4+\l,1);
	\draw[ white, double = black, line width = 2pt, double distance = 1pt ] 
		(1,4,3) -- (3,4,3) -- (3,4,1) -- (3,4+\l,1) -- (1,4+\l,1) -- (1,4+\l,3) --  cycle;
	\begin{scope}
	\clip (1,4,1) circle (.5cm)
	(3,4,1) circle (.5cm)
	(3,4,3) circle (.5cm)
	(1,4,3) circle (.5cm)
	(1,4+\l,1) circle (.5cm)
	(3,4+\l,1) circle (.5cm)
	(3,4+\l,3) circle (.5cm)
	(1,4+\l,3) circle (.5cm);	
	\draw[ black, line width = 1pt] 	
		(1,4,1) -- (3,4,1)
		(1,4,1) -- (1,4,3)
		(1,4,1) -- (1,4+\l,1);
	\draw[ black, line width = 1pt] 	
		(3,4+\l,3) -- (3,4,3)
		(3,4+\l,3) -- (1,4+\l,3)
		(3,4+\l,3) -- (3,4+\l,1);
	\draw[ black, line width = 1pt] 	
		(1,4,3) -- (3,4,3) -- (3,4,1) -- (3,4+\l,1) -- (1,4+\l,1) -- (1,4+\l,3) --  cycle;
	\end{scope}
	
	\draw[ white, double = black, line width = 2pt, double distance = 1pt ] 
		(0,4+\l,0) -- (0,4+\l,4) 
		(0,4+\l,0) -- (0,8+\l,0)
		(0,4+\l,0) -- (4,4+\l,0);	
	\draw[ white, double = black, line width = 2pt, double distance = 1pt ] 
		(4,8+\l,4) -- (4,8+\l,0)
		(4,8+\l,4) -- (4,4+\l,4)
		(4,8+\l,4) -- (0,8+\l,4);
	\draw[ white, double = black, line width = 2pt, double distance = 1pt ] 
		(4,8+\l,0) -- (4,4+\l,0) -- (4,4+\l,4) -- (0,4+\l,4) -- (0,8+\l,4) -- (0,8+\l,0) -- cycle;
	\begin{scope}
	\clip (0,4+\l,0) circle (.5cm)
	(4,4+\l,0) circle (.5cm)
	(0,4+\l,4) circle (.5cm)
	(4,4+\l,4) circle (.5cm)
	(0,8+\l,0) circle (.5cm)
	(4,8+\l,0) circle (.5cm)
	(0,8+\l,4) circle (.5cm)
	(4,8+\l,4) circle (.5cm);	
	\draw[ black, line width = 1pt] 	
		(0,4+\l,0) -- (0,4+\l,4) 
		(0,4+\l,0) -- (0,8+\l,0)
		(0,4+\l,0) -- (4,4+\l,0);	
	\draw[ black, line width = 1pt] 	
		(4,8+\l,4) -- (4,8+\l,0)
		(4,8+\l,4) -- (4,4+\l,4)
		(4,8+\l,4) -- (0,8+\l,4);
	\draw[ black, line width = 1pt] 	
		(4,8+\l,0) -- (4,4+\l,0) -- (4,4+\l,4) -- (0,4+\l,4) -- (0,8+\l,4) -- (0,8+\l,0) -- cycle;
	\end{scope}
	
  	\draw[draw=darkred,line width=.5pt,|-|]    (4+\dx,0-\dx,0) to  (4+\dx,0-\dx,1-\de);
  	\draw[draw=darkred,line width=.5pt,|-|]    (4+\dx,0-\dx,1+\de) to  (4+\dx,0-\dx,3-\de);
  	\draw[draw=darkred,line width=.5pt,|-|]    (4+\dx,0-\dx,3+\de) to  (4+\dx,0-\dx,4);
   	\draw[draw=darkred,line width=.5pt,|-|]    (4+\dx,0,0-\dx) to  (4+\dx,4-\de,0-\dx);
   	\draw[draw=darkred,line width=.5pt,|-|]    (4+\dx,4+\de,0-\dx) to  (4+\dx,4+\l-\de,0-\dx);
   	\draw[draw=darkred,line width=.5pt,|-|]    (4+\dx,4+\l+\de,0-\dx) to  (4+\dx,8+\l,0-\dx);
   	\draw[draw=darkred,line width=.5pt,|-|]    (4,8+\l+\dx,0-\dx) to  (3+\de,8+\l+\dx,0-\dx);
   	\draw[draw=darkred,line width=.5pt,|-|]    (3-\de,8+\l+\dx,0-\dx) to  (1+\de,8+\l+\dx,0-\dx);
   	\draw[draw=darkred,line width=.5pt,|-|]    (1-\de,8+\l+\dx,0-\dx) to  (0,8+\l+\dx,0-\dx);
   	
	\node at (.5,8+\l+\dl,0-\dl){\red{$l_x$}};
	\node at (2,8+\l+\dl,0-\dl){\red{$l_x'$}};
	\node at (3.5,8+\l+\dl,0-\dl){\red{$l_x''$}};
	\node at (4+\dl,6+\l,0-\dl){\red{$l_y$}};
	\node at (4+\dl,4+\l/2,0-\dl){\red{$l_y'$}};
	\node at (4+\dl,2,0-\dl){\red{$l_y$}};
	\node at  (4+\dl,0-\dl,.5){\red{$l_z$}};
	\node at (4+\dl,0-\dl,2){\red{$l_z'$}};
	\node at (4+\dl,0-\dl,3.5){\red{$l_z''$}};
	
	\node at (2,0,2) {\Large $A$};
	\node at (2,4+\l/2,2) {\large $B$};
	\node at (2,8+\l,2) {\Large $C$};

	\end{tikzpicture}
	}
\end{subfigure}
 \caption{ 
 Regions used by Ref.~\onlinecite{PhysRevB.97.125101} (upper left), and Ref.~\onlinecite{Grover2011} (lower left), to calculate the topological entanglement entropy in 3D, and used here (right) to define the dumbell entropy in 3D. }
\label{3Dentropyregions}
\end{figure}

In three or more spatial dimensions it is a more subtle issue to define the topological entanglement entropy~\cite{Grover2011}. 
In Fig.~\ref{3Dentropyregions}~(left) we depict regions used by Refs.~\onlinecite{PhysRevB.97.125101,Grover2011} to calculate TEE that are `valid' in the terminology of Ref.~\onlinecite{Grover2011}. We remark that similar results hold for the other regions considered in Refs.~\onlinecite{PhysRevB.97.125101,Grover2011}, although the other region considered in Ref.~\onlinecite{PhysRevB.97.125101} is not `valid'. Due to the possibility of fracton topological order in 3D, even `valid' configurations for the TEE may not lead to constant, topologically invariant contributions~\cite{PhysRevB.97.125101}. 

Here we demonstrate that SSPTs may lead to linear and constant corrections to the area law, thus making it tricky to extract information about a fracton topological phase unambiguously from either linear or constant terms in the TEE as was suggested in Ref.~\onlinecite{PhysRevB.97.125101}. In particular we show that a trivial  phase may display a linear and constant term in the TEE due to nontrivial SSPT order.  For this reason we generalize the dumbell entropy to 3D, see Fig.~\ref{3Dentropyregions}~(right), to capture spurious contributions to the TEE due to linear subsystem symmetries. We remark that it is unlikely that planar, or fractal subsystem symmetries will lead to spurious  TEE as they do not fit into the regions commonly used to calculate the TEE. 

The 3D cluster model on the body centered cubic lattice is given by translates of the following stabilizer generators~\cite{you2018subsystem} 
\begin{align}
\drawgenerator{ZI}{ZI}{ZI}{ZI}{ZI}{ZX}{ZI}{ZI}
&&
\drawgenerator{IZ}{XZ}{IZ}{IZ}{IZ}{IZ}{IZ}{IZ}
\end{align}
This model has linear $\Z_2 \times \Z_2$ subsystem symmetries generated by a tensor product of $XI$ or $IX$ along a line in the $x,\,y,\,$ or $z$ direction.  This is a direct generalization of the 2D square lattice cluster model, and it has a clear generalization to any higher dimensional hypercubic lattice. 

The topological entanglement entropy of the 3D cluster model for each of the regions shown in Fig.~\ref{3Dentropyregions} is given by
\begin{align}
S_{\text{topo}}^{\text{3D}} &= - 2 l_z + 2
\, , \\
S_{\text{topo}}^{\text{3D}'} &= - 2 l_z + 2
\, , \\
S_{\text{dumb}}^{\text{3D}} &= - 2 l_x' - 2 l_z' +2
\, , 
\end{align}
for the regions depicted in the upper left, lower left, and right of Fig.~\ref{3Dentropyregions}, respectively. The above calculation assumes none of the lengths defined in Fig.~\ref{3Dentropyregions}  are zero.

\section{String correlation length of deformed cluster states}
\label{sec:numerics}

We have looked at a family of short-range entangled states $\ket{\theta}$ that interpolates between the $\ket{+}^{\otimes N}$ state, for $\theta=0$, and the cluster state, for $\theta=\pi$. These states are produced by local unitary circuits as follows
\begin{align}
\ket{\theta} := \prod_{\langle i , j \rangle} {C}\theta_{i,j} \ket{+}^{\otimes N}
\, ,
\end{align}
where $\langle i , j \rangle$ denotes qubits that share an edge from the original square lattice, and 
\begin{align}
 {C}\theta =  
 \begin{pmatrix}
 1 & 0 & 0 & 0 \\
0  & 1 & 0 & 0  \\
0  & 0 & 1 & 0 \\
0  & 0 & 0 & e^{i\theta}
 \end{pmatrix}
\, .
\end{align}

This leads to a natural tensor network representation of the state $\ket{\theta}$, found by contracting 
\begin{align}
	\begin{array}{c}
	\includeTikz{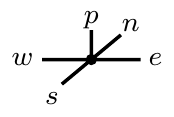}{
	\begin{tikzpicture}[scale=.1]
	\draw[draw=black,line width=1pt] (-5,0) -- (0,0) -- (5,0) 
	(-3,-2.5) -- (0,0) -- (3,2.5)
	(0,0) -- (0,3);
 	\draw[color=black,fill=black]  (0,0) circle (.5);
	\node at (0,4){\footnotesize $p$};
	\node at (4,3.5){\footnotesize $n$};
	\node at (-4,-4){\footnotesize $s$};
	\node at (-7,0){\footnotesize $w$};
	\node at (6.5,0){\footnotesize $e$};
	\end{tikzpicture}
	}
	\end{array}
	= 
	\frac{1}{\sqrt{2}}\delta_{p=n=e=w=s}
\, ,
\end{align}
 tensors on vertices with 
 \begin{align}
	\begin{array}{c}
	\includeTikz{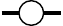}{
	\begin{tikzpicture}[scale=.1]
	\draw[draw=black,line width=1pt]
	(-3,0) -- (3,0);
	\draw[color=black,fill=white] (0,0) circle (1.25);
	\end{tikzpicture}
	}
	\end{array}
	 = 
 \begin{pmatrix}
1 & 1 \\
1 & e^{i\theta}
\end{pmatrix}
 \end{align}
  matrices on edges. 
  
  It is known from the study of SPT phases in one dimension~\cite{pollmann2010entanglement} that the gap of the entanglement spectrum $\lambda_1 - \lambda_2$ serves as an order parameter, obtaining the value $0$ for nontrivial SPT phases. It is also known that long-range string order is a characteristic property of nontrivial SPT phases protected by an on-site unitary symmetry. Hence the string correlation length should diverge with $- \log (\lambda_1 - \lambda_2) $ as a nontrivial SPT phase is approached. 
A rescaled reduced density matrix  $\rho/ \sqrt{\tr \rho^2}$  from a subregion of an SSPT state that supports a one dimensional symmetry can be thought of as a properly normalized one dimensional SPT state. 
Hence we can use the entanglement spectrum of such a density matrix to probe the string correlation length related to SSPT order. 

\begin{figure}[t]
\center
\begin{subfigure}{0.49\linewidth}
 \includegraphics[scale=0.5]{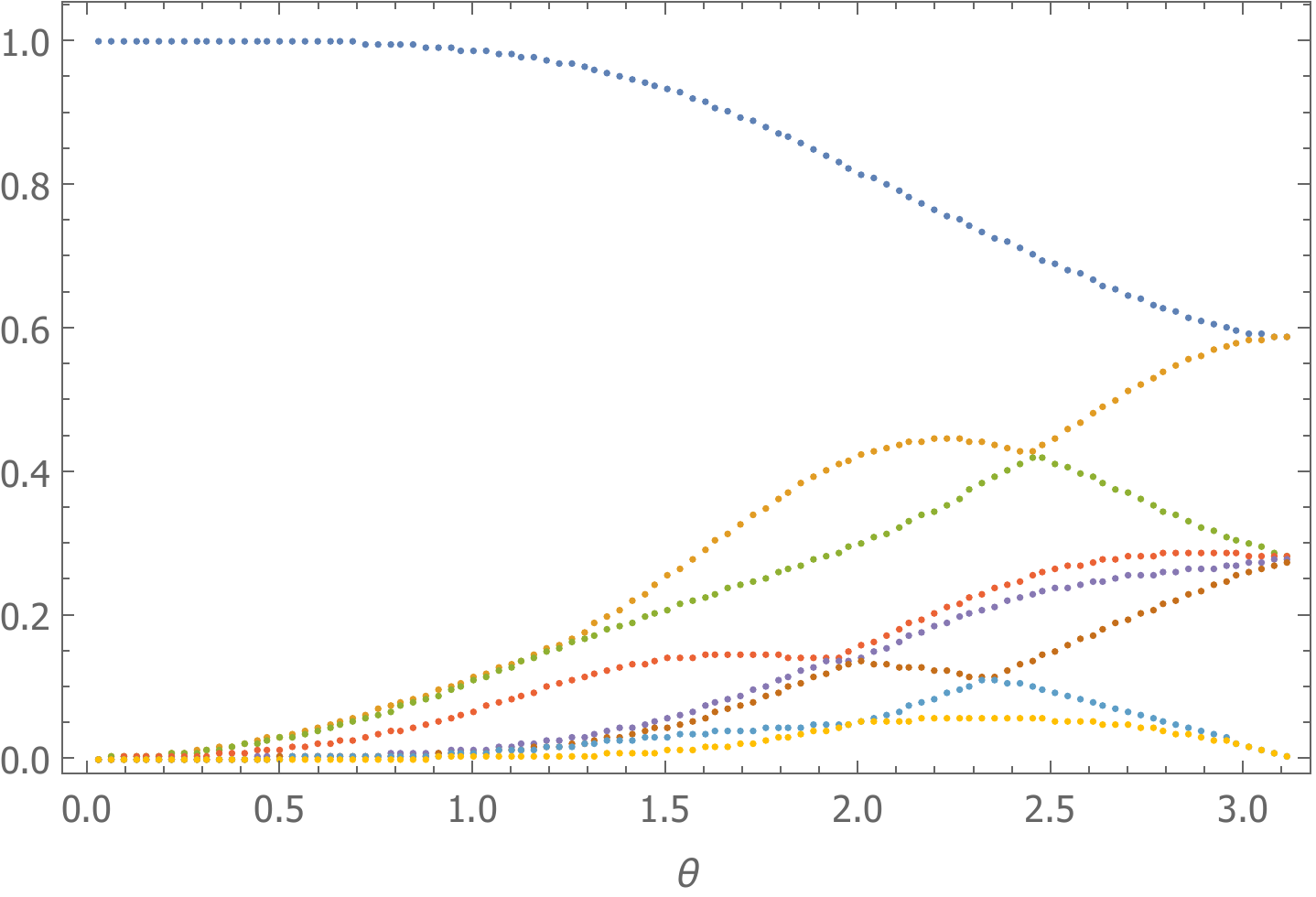} 
\end{subfigure}
\begin{subfigure}{0.49\linewidth}
 \includegraphics[scale=0.5]{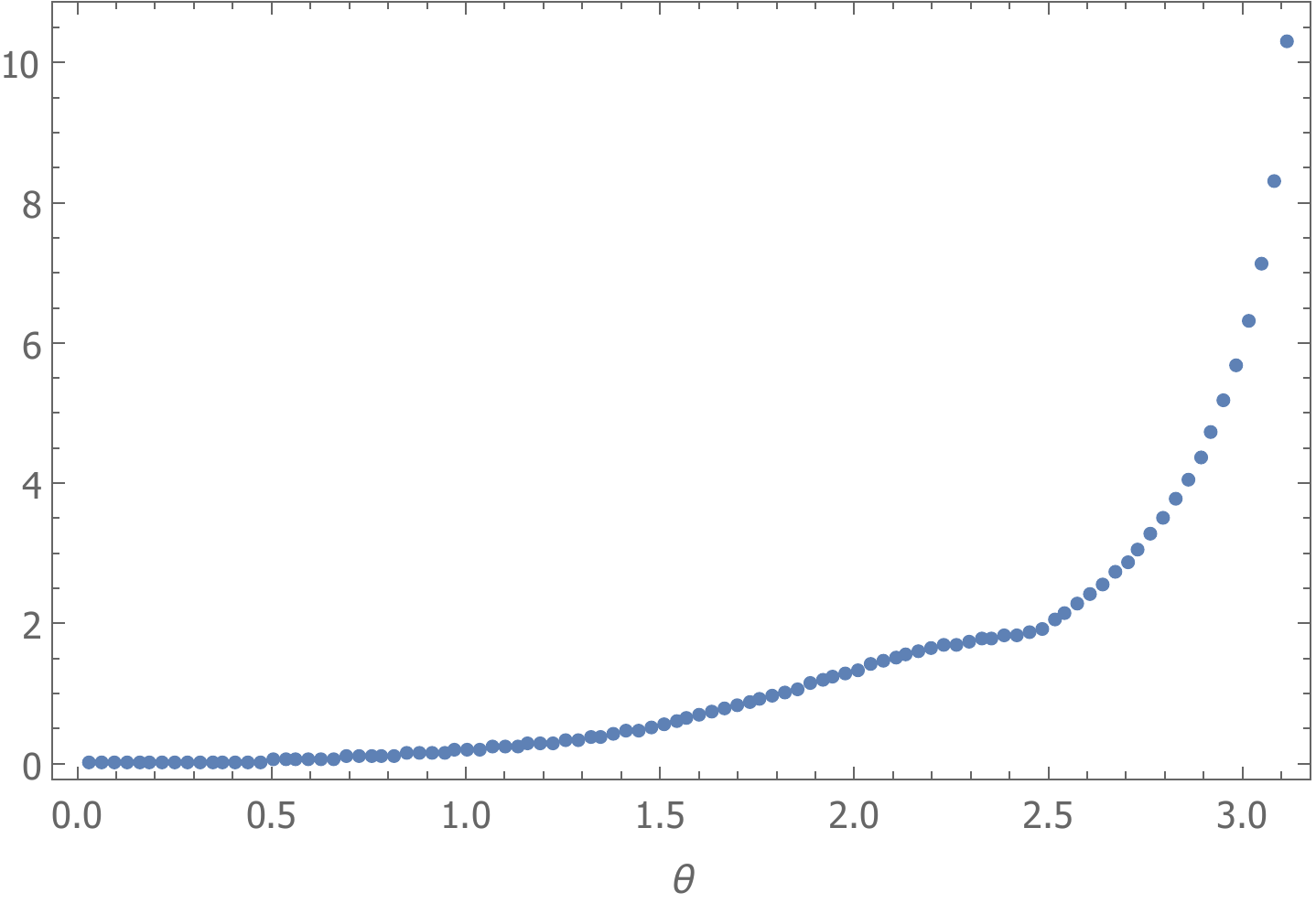}
 \end{subfigure}
 \caption{ (left) Leading Schmidt values $\lambda_i$ as functions of $\theta$.  (right) $- \log (\lambda_1-\lambda_2) $ as a function of $\theta$. }
\label{Espectheta}
\end{figure}
The entanglement spectrum in Fig.~\ref{Espectheta}~(left) was obtained by considering the reduced density matrix of $\ket{\theta}$ on a single row $r$ of the lattice
\begin{align}
\rho_\theta = {\tr}_{(i,j\neq r)} \ket{\theta} \bra{\theta}
\, .
\end{align}
All the C$\theta$ gates that are not adjacent to row $r$ cancel in this trace, leaving a low bond dimension matrix product operator (MPO) with local tensor
\begin{align}
2^{-4}
	\begin{array}{c}
	\includeTikz{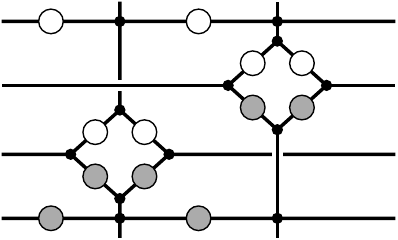}{
	\begin{tikzpicture}[scale=.1]
	\def\y{4.5}
	\def\x{5}
	\def\e{12}
	\def\eh{5}
	\def\ep{2}
	\def\h{20}
	\def\w{40}
	\def\r{1.25}
	\def\d{.5}
	\draw[draw=black,line width=1pt] (0,0) -- (\w,0) 
	 (0,\h) -- (\w,\h) 
	 (\e,0+\ep) -- (\e+\x,\y+\ep) -- (\e,2*\y+\ep) -- (\e-\x,\y+\ep) -- cycle
 	 (\w-\e,\h-\ep) -- (\w-\e+\x,\h-\y-\ep) -- (\w-\e,\h-2*\y-\ep) -- (\w-\e-\x,\h-\y-\ep) -- cycle
 	 (\e,2*\y+\ep) -- (\e,\h+\ep) 
 	 (\e,+\ep) -- (\e,-\ep)
 	(\e+\x,\y+\ep) -- (\w,\y+\ep)
	(\e-\x,\y+\ep) -- (0,\y+\ep);
	\draw[draw=white,line width=3pt] 	(\e-\ep,\h-\y-\ep) -- (\e+\ep,\h-\y-\ep)
	(\w-\e,\y) -- (\w-\e,\y+2*\ep)
	 ;
	\draw[draw=black,line width=1pt] (\w-\e,\h-\ep) -- (\w-\e,\h+\ep)
	(\w-\e,\h-2*\y-\ep) -- (\w-\e,-\ep)
	(\w-\e+\x,\h-\y-\ep) -- (\w,\h-\y-\ep)
	(\w-\e-\x,\h-\y-\ep) -- (0,\h-\y-\ep) 
	 ;
 	\draw[color=black,fill=black] (\e,0) circle (\d);
 	\draw[color=black,fill=black] (\e,\h) circle (\d);
 	\draw[color=black,fill=black] (\w-\e,0) circle (\d);
 	\draw[color=black,fill=black] (\w-\e,\h) circle (\d);
 	\draw[color=black,fill=black] (\e,0+\ep) circle (\d);
 	\draw[color=black,fill=black] (\e+\x,\y+\ep) circle (\d);
 	\draw[color=black,fill=black] (\e,2*\y+\ep) circle (\d);
 	\draw[color=black,fill=black] (\e-\x,\y+\ep) circle (\d);
 	\draw[color=black,fill=black] (\w-\e,\h-\ep) circle (\d);
 	\draw[color=black,fill=black] (\w-\e+\x,\h-\y-\ep) circle (\d);
 	\draw[color=black,fill=black]  (\w-\e,\h-2*\y-\ep) circle (\d);
 	\draw[color=black,fill=black]  (\w-\e-\x,\h-\y-\ep) circle (\d);
	\draw[color=black,fill=black!33] (\eh,0) circle (\r);
	\draw[color=black,fill=black!33] (\w/2,0) circle (\r);
	\draw[color=black,fill=black!33]  (\e+\x/2,\y/2+\ep)  circle (\r);
	\draw[color=black,fill=black!33]  (\e-\x/2,\y/2+\ep)  circle (\r);
	\draw[color=black,fill=black!33]   (\w-\e+\x/2,\h-1.5*\y-\ep) circle (\r);
	\draw[color=black,fill=black!33]   (\w-\e-\x/2,\h-1.5*\y-\ep) circle (\r);
	\draw[color=black,fill=white] (\eh,\h) circle (\r);
	\draw[color=black,fill=white] (\w/2,\h) circle (\r);
	\draw[color=black,fill=white]  (\e+\x/2,1.5*\y+\ep)  circle (\r);
	\draw[color=black,fill=white]  (\e-\x/2,1.5*\y+\ep)  circle (\r);
	\draw[color=black,fill=white]   (\w-\e+\x/2,\h-.5*\y-\ep) circle (\r);
	\draw[color=black,fill=white]   (\w-\e-\x/2,\h-.5*\y-\ep) circle (\r);
	\end{tikzpicture}
	}
	\end{array}
\, ,
\end{align}
where 
\begin{align}
	\begin{array}{c}
	\includeTikz{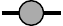}{
	\begin{tikzpicture}[scale=.1]
	\draw[draw=black,line width=1pt]
	(-3,0) -- (3,0);
	\draw[color=black,fill=black!33] (0,0) circle (1.25);
	\end{tikzpicture}
	}
	\end{array}
	=
	 \begin{pmatrix}
1 & 1 \\
1 & e^{-i\theta}
\end{pmatrix}
\, .
\end{align}
We perform the rescaling $\rho_\theta/ \sqrt{\tr \rho_\theta^2}$ such that the resulting MPO can be thought of as a one dimensional pure state $\ket{\rho_\theta}$. 
Next we perform a singular value decomposition of the one dimensional state with respect to a bipartition across some cut
\begin{align}
 \ket{\rho_\theta}  = \sum_{s } \lambda_s  \ket{L_s } \ket{R_s}
\, ,
\end{align}
where the set of states $L_s$ and $R_s$ are orthogonal. 
This can be found sequentially by singular value (or $RQ$) decomposing the MPO tensors to the right of the cut, starting at $\infty$, into a square matrix times an isometry, the former of which can be absorbed into the next tensor to the left along the chain
\begin{align}
\begin{array}{c}
	\includeTikz{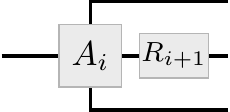}{
	\begin{tikzpicture}[scale=.1]
	\def\x{14}
	\def\y{5.5}
	\def\s{8.5}
	\draw[draw=black,line width=1pt]  (-\x+5,0) -- (\x,0) 
	(\x,-\y) -- (0,-\y) -- (0,\y) --(\x,\y)
;
	\node[transform shape,draw=black!29,fill=black!10!white!80,regular polygon,regular polygon sides=4,minimum size=9cm] at (0,0){};
	\node at (0,0) { $A_i$};
	\node[transform shape,draw=black!29,fill=black!10!white!80,minimum height=4.5cm,minimum width=7cm] at (\s,0){};
	\node at (\s,0) {\footnotesize $R_{i+1}$};
	\end{tikzpicture}
	}
	\end{array}
	=
\begin{array}{c}
	\includeTikz{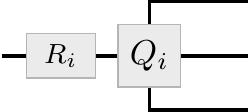}{
	\begin{tikzpicture}[scale=.1]
	\def\x{10}
	\def\y{5.5}
	\def\s{9}
	\draw[draw=black,line width=1pt]  (-\x-5,0) -- (\x,0) 
	(\x,-\y) -- (0,-\y) -- (0,\y) --(\x,\y)
;
	\node[transform shape,draw=black!29,fill=black!10!white!80,regular polygon,regular polygon sides=4,minimum size=9cm] at (0,0){};
	\node at (0,0) { $Q_i$};
	\node[transform shape,draw=black!29,fill=black!10!white!80,minimum height=4.5cm,minimum width=7cm] at (-\s,0){};
	\node at (-\s,0) {\footnotesize $R_{i}$};
	\end{tikzpicture}
	}
	\end{array}
\, .
\end{align}
 A similar process is carried out for MPO tensors to the left of the cut, and this is run until convergence. The results are shown in Fig.~\ref{Espectheta}~(left). We have also plotted $- \log (\lambda_1 - \lambda_2) $ in Fig.~\ref{Espectheta}~(right).

Similar to the above, when calculating the entropy of the reduced density matrix of $\ket{\theta}$ on some region $R$, only gates crossing the boundary $\partial R$ contribute. This is because all gates that act completely within or completely outside $R$ will cancel.  
For any integer R\`enyi entropy $S^{(n)}$ with $n>1$, see Eq.~\eqref{eq:renyientropy}, the calculation of  $S^{(n)}(\rho_R)$ can be reduced to a multiplication of matrices along the boundary $\partial R$. Here we focus on the case $n=2$ for simplicity, as the result should not qualitatively depend on $n$ anyway~\cite{topologicalrenyi,PhysRevA.86.062330}. 
  From the tensor network representation of $\ket{\theta}$ one can derive the matrix that occurs for each unit length along a horizontal or vertical boundary
  \begin{align}
  \mathbb{E}
  := 2^{-4}
  	\begin{array}{c}
	\includeTikz{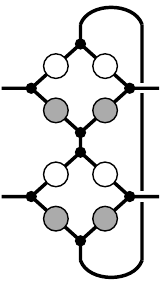}{
	\begin{tikzpicture}[scale=.1]
	\def\y{4.5}
	\def\x{5}
	\def\e{8}
	\def\eh{5}
	\def\ep{2}
	\def\h{20}
	\def\w{40}
	\def\r{1.25}
	\def\d{.5}
	\def\l{1.25}
	\draw[draw=black,line width=1pt] 
 	 (\w-\e,\h-\ep) -- (\w-\e+\x,\h-\y-\ep) -- (\w-\e,\h-2*\y-\ep) -- (\w-\e-\x,\h-\y-\ep) -- cycle
 	 (\w-\e,\h-\ep-2*\y-\ep) -- (\w-\e+\x,\h-\y-\ep-2*\y-\ep) -- (\w-\e,\h-2*\y-\ep-2*\y-\ep) -- (\w-\e-\x,\h-\y-\ep-2*\y-\ep) -- cycle
 	 (\w-\e+\x+\l,\h) --  (\w-\e+\x+\l,\h-2*\y-2*\ep-2*\y-\ep)
	;
	\draw[draw=white,line width=3pt]	(\w-\e+\x,\h-\y-\ep) -- (\w,\h-\y-\ep)
	(\w-\e+\x,\h-\y-\ep-2*\y-\ep) -- (\w,\h-\y-\ep-2*\y-\ep)
	 ;
	\draw[draw=black,line width=1pt] (\w-\e,\h-\ep) -- (\w-\e,\h)
	(\w-\e,\h-2*\y-\ep) -- (\w-\e,\h-2*\y-2*\ep)
	(\w-\e+\x,\h-\y-\ep) -- (\w,\h-\y-\ep)
	(\w-\e-\x,\h-\y-\ep) -- (\w-2*\e,\h-\y-\ep) 
	(\w-\e,\h-\ep-2*\y-\ep) -- (\w-\e,\h-2*\y-\ep)
	(\w-\e,\h-2*\y-\ep-2*\y-\ep) -- (\w-\e,\h-2*\y-2*\ep-2*\y-\ep)
	(\w-\e+\x,\h-\y-\ep-2*\y-\ep) -- (\w,\h-\y-\ep-2*\y-\ep)
	(\w-\e-\x,\h-\y-\ep-2*\y-\ep) -- (\w-2*\e,\h-\y-\ep-2*\y-\ep) 
	(\w-\e,\h) to [bend left=70]  (\w-\e+\x+\l,\h) 
	 (\w-\e,\h-2*\y-2*\ep-2*\y-\ep) to [bend right=70] (\w-\e+\x+\l,\h-2*\y-2*\ep-2*\y-\ep)
	 ;
 	\draw[color=black,fill=black] (\w-\e,\h-\ep) circle (\d);
 	\draw[color=black,fill=black] (\w-\e+\x,\h-\y-\ep) circle (\d);
 	\draw[color=black,fill=black]  (\w-\e,\h-2*\y-\ep) circle (\d);
 	\draw[color=black,fill=black]  (\w-\e-\x,\h-\y-\ep) circle (\d);
 	\draw[color=black,fill=black] (\w-\e,\h-\ep-2*\y-\ep) circle (\d);
 	\draw[color=black,fill=black] (\w-\e+\x,\h-\y-\ep-2*\y-\ep) circle (\d);
 	\draw[color=black,fill=black]  (\w-\e,\h-2*\y-\ep-2*\y-\ep) circle (\d);
 	\draw[color=black,fill=black]  (\w-\e-\x,\h-\y-\ep-2*\y-\ep) circle (\d);
	\draw[color=black,fill=black!33]   (\w-\e+\x/2,\h-1.5*\y-\ep) circle (\r);
	\draw[color=black,fill=black!33]   (\w-\e-\x/2,\h-1.5*\y-\ep) circle (\r);
	\draw[color=black,fill=white]   (\w-\e+\x/2,\h-.5*\y-\ep) circle (\r);
	\draw[color=black,fill=white]   (\w-\e-\x/2,\h-.5*\y-\ep) circle (\r);
	\draw[color=black,fill=black!33]   (\w-\e+\x/2,\h-1.5*\y-\ep-2*\y-\ep) circle (\r);
	\draw[color=black,fill=black!33]   (\w-\e-\x/2,\h-1.5*\y-\ep-2*\y-\ep) circle (\r);
	\draw[color=black,fill=white]   (\w-\e+\x/2,\h-.5*\y-\ep-2*\y-\ep) circle (\r);
	\draw[color=black,fill=white]   (\w-\e-\x/2,\h-.5*\y-\ep-2*\y-\ep) circle (\r);
	\end{tikzpicture}
	}
	\end{array}
  \, ,
  \end{align}
  and for corners shown in Fig.~\ref{entropyregion} 
   \begin{align}
C
  &:= 2^{-6}
  	\begin{array}{c}
	\includeTikz{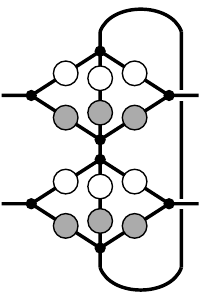}{
	\begin{tikzpicture}[scale=.1]
	\def\y{4.5}
	\def\x{7}
	\def\e{10}
	\def\eh{5}
	\def\ep{2}
	\def\h{20}
	\def\w{40}
	\def\r{1.25}
	\def\d{.5}
	\def\l{1.25}
	\draw[draw=black,line width=1pt] 
 	 (\w-\e,\h-\ep) -- (\w-\e+\x,\h-\y-\ep) -- (\w-\e,\h-2*\y-\ep) -- (\w-\e-\x,\h-\y-\ep) -- cycle
	(\w-\e,\h-\ep) -- (\w-\e,\h-2*\y-\ep) 
 	 (\w-\e,\h-\ep-2*\y-\ep) -- (\w-\e+\x,\h-\y-\ep-2*\y-\ep) -- (\w-\e,\h-2*\y-\ep-2*\y-\ep) -- (\w-\e-\x,\h-\y-\ep-2*\y-\ep) -- cycle
	(\w-\e,\h-\ep-2*\y-\ep)  -- (\w-\e,\h-2*\y-\ep-2*\y-\ep) 
 	 (\w-\e+\x+\l,\h) --  (\w-\e+\x+\l,\h-2*\y-2*\ep-2*\y-\ep)
	;
	\draw[draw=white,line width=3pt]	(\w-\e+\x,\h-\y-\ep) -- (\w,\h-\y-\ep)
	(\w-\e+\x,\h-\y-\ep-2*\y-\ep) -- (\w,\h-\y-\ep-2*\y-\ep)
	 ;
	\draw[draw=black,line width=1pt] (\w-\e,\h-\ep) -- (\w-\e,\h)
	(\w-\e,\h-2*\y-\ep) -- (\w-\e,\h-2*\y-2*\ep)
	(\w-\e+\x,\h-\y-\ep) -- (\w,\h-\y-\ep)
	(\w-\e-\x,\h-\y-\ep) -- (\w-2*\e,\h-\y-\ep) 
	(\w-\e,\h-\ep-2*\y-\ep) -- (\w-\e,\h-2*\y-\ep)
	(\w-\e,\h-2*\y-\ep-2*\y-\ep) -- (\w-\e,\h-2*\y-2*\ep-2*\y-\ep)
	(\w-\e+\x,\h-\y-\ep-2*\y-\ep) -- (\w,\h-\y-\ep-2*\y-\ep)
	(\w-\e-\x,\h-\y-\ep-2*\y-\ep) -- (\w-2*\e,\h-\y-\ep-2*\y-\ep) 
	(\w-\e,\h) to [bend left=70]  (\w-\e+\x+\l,\h) 
	 (\w-\e,\h-2*\y-2*\ep-2*\y-\ep) to [bend right=70] (\w-\e+\x+\l,\h-2*\y-2*\ep-2*\y-\ep)
	 ;
 	\draw[color=black,fill=black] (\w-\e,\h-\ep) circle (\d);
 	\draw[color=black,fill=black] (\w-\e+\x,\h-\y-\ep) circle (\d);
 	\draw[color=black,fill=black]  (\w-\e,\h-2*\y-\ep) circle (\d);
 	\draw[color=black,fill=black]  (\w-\e-\x,\h-\y-\ep) circle (\d);
 	\draw[color=black,fill=black] (\w-\e,\h-\ep-2*\y-\ep) circle (\d);
 	\draw[color=black,fill=black] (\w-\e+\x,\h-\y-\ep-2*\y-\ep) circle (\d);
 	\draw[color=black,fill=black]  (\w-\e,\h-2*\y-\ep-2*\y-\ep) circle (\d);
 	\draw[color=black,fill=black]  (\w-\e-\x,\h-\y-\ep-2*\y-\ep) circle (\d);
	\draw[color=black,fill=black!33]   (\w-\e+\x/2,\h-1.5*\y-\ep) circle (\r);
	\draw[color=black,fill=black!33]   (\w-\e-\x/2,\h-1.5*\y-\ep) circle (\r);
	\draw[color=black,fill=white]   (\w-\e+\x/2,\h-.5*\y-\ep) circle (\r);
	\draw[color=black,fill=white]   (\w-\e-\x/2,\h-.5*\y-\ep) circle (\r);
	\draw[color=black,fill=black!33]   (\w-\e+\x/2,\h-1.5*\y-\ep-2*\y-\ep) circle (\r);
	\draw[color=black,fill=black!33]   (\w-\e-\x/2,\h-1.5*\y-\ep-2*\y-\ep) circle (\r);
	\draw[color=black,fill=white]   (\w-\e+\x/2,\h-.5*\y-\ep-2*\y-\ep) circle (\r);
	\draw[color=black,fill=white]   (\w-\e-\x/2,\h-.5*\y-\ep-2*\y-\ep) circle (\r);
	\draw[color=black,fill=black!33]   (\w-\e,\h-1.5*\y-\ep+.5) circle (\r);
	\draw[color=black,fill=white]   (\w-\e,\h-.5*\y-\ep-.5) circle (\r);
	\draw[color=black,fill=black!33]   (\w-\e,\h-1.5*\y-\ep-2*\y-\ep+.5) circle (\r);
	\draw[color=black,fill=white]   (\w-\e,\h-.5*\y-\ep-2*\y-\ep-.5) circle (\r);
	\end{tikzpicture}
	}
	\end{array}
  \, ,  
  \\
\ket{R} &:= C \, ( \ket{+} \otimes \ket{+} )
\, ,  \\
\bra{L} &:= (\ket{R})^T
  \, ,
  \\
C_\Gamma &:= 
 \begin{pmatrix}
 1 & 0 & 0 & 0 \\
0  & |\frac{(1+e^{i \theta})}{2}|^2 & 0 & 0  \\
0  & 0 & |\frac{(1+e^{i \theta})}{2}|^2  & 0 \\
0  & 0 & 0 & 1
 \end{pmatrix}
\, ,
\\
\ket{R_\Gamma} &:=  C_\Gamma \, ( \ket{+} \otimes \ket{+} )
\, ,
\\
 \bra{L_\Gamma} &:= (\ket{R_\Gamma})^T
\, .
  \end{align}

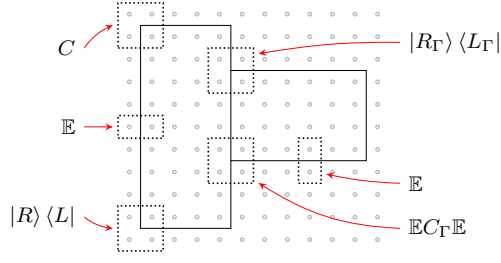
\begin{figure}[t]
\center
	\includeTikz{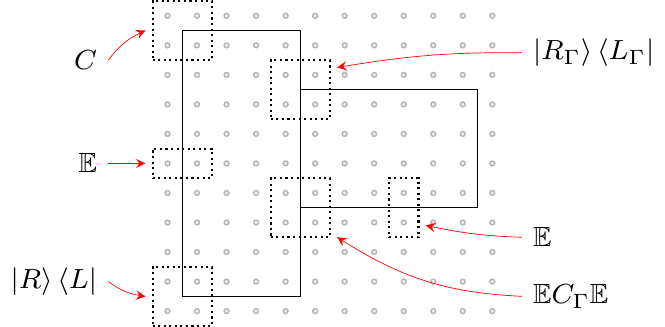}{
	\begin{tikzpicture}[scale=.1]
\def\dy{3}
\def\w{.25}
\def\n{10}
\def\eps{.25}
			\foreach \x in {0,1,...,11} {
        \foreach \y in {0,1,...,\n} {
		  \draw[color=black!29,fill=black!10!white!80] (\x*\dy,\y*\dy) circle (\w);
        }
        }
	\draw[draw=black,line width=.2pt] (.5*\dy,.5*\dy) -- (4.5*\dy,.5*\dy) --  (4.5*\dy,9.5*\dy) -- (.5*\dy,9.5*\dy) -- cycle 
 	 (4.5*\dy,3.5*\dy) --  (10.5*\dy,3.5*\dy) -- (10.5*\dy,7.5*\dy) --  (4.5*\dy,7.5*\dy);
	\draw[draw=black,line width=.6pt,densely dotted] 
	(-.5*\dy,-.5*\dy) -- (1.5*\dy,-.5*\dy) -- (1.5*\dy,1.5*\dy) -- (-.5*\dy,1.5*\dy) -- cycle 
	(-.5*\dy,8.5*\dy) -- (1.5*\dy,8.5*\dy) -- (1.5*\dy,10.5*\dy) -- (-.5*\dy,10.5*\dy) -- cycle 
	(-.5*\dy,4.5*\dy) -- (1.5*\dy,4.5*\dy) -- (1.5*\dy,5.5*\dy) -- (-.5*\dy,5.5*\dy) -- cycle 
	(3.5*\dy,6.5*\dy) -- (5.5*\dy,6.5*\dy) -- (5.5*\dy,8.5*\dy) -- (3.5*\dy,8.5*\dy) -- cycle 
	(3.5*\dy,2.5*\dy) -- (5.5*\dy,2.5*\dy) -- (5.5*\dy,4.5*\dy) -- (3.5*\dy,4.5*\dy) -- cycle 
	(7.5*\dy,2.5*\dy) -- (8.5*\dy,2.5*\dy) -- (8.5*\dy,4.5*\dy) -- (7.5*\dy,4.5*\dy) -- cycle 
	;
	\draw[draw=red,line width=.2pt,decoration={markings,mark=at position 1 with {\arrow[scale=1,red]{stealth}}},
    postaction={decorate}]     (12*\dy,8.75*\dy) to [bend right=5] ( 5.5*\dy+\eps*\dy , 8.25*\dy);
	\draw[draw=red,line width=.2pt,decoration={markings,mark=at position 1 with {\arrow[scale=1,red]{stealth}}},
    postaction={decorate}]   (12*\dy,2.5*\dy) to [bend left=5] ( 8.5*\dy +\eps*\dy , 2.9*\dy);
    	\draw[draw=red,line width=.2pt,decoration={markings,mark=at position 1 with {\arrow[scale=1,red]{stealth}}},
    postaction={decorate}]   (12*\dy,.5*\dy) to [bend left=15] ( 5.5*\dy +\eps*\dy , 2.5*\dy);
    	\draw[draw=red,line width=.2pt,decoration={markings,mark=at position 1 with {\arrow[scale=1,red]{stealth}}},
    postaction={decorate}]    (-2*\dy,8.5*\dy) to [bend left=15] ( -.5*\dy-\eps*\dy , 9.5*\dy);
   	\draw[draw=red,line width=.2pt,decoration={markings,mark=at position 1 with {\arrow[scale=1,red]{stealth}}},
    postaction={decorate}]    (-2*\dy,1*\dy) to [bend right=15] ( -.5*\dy -\eps*\dy, .5*\dy);
    	\draw[draw=red,line width=.2pt,decoration={markings,mark=at position 1 with {\arrow[scale=1,red]{stealth}}},
    postaction={decorate}]    (-2*\dy,5*\dy) to ( -.5*\dy-\eps*\dy, 5*\dy);
	\node[anchor=west] at (12*\dy,8.75*\dy) {\footnotesize $\ket{R_\Gamma}\bra{L_\Gamma}$};
	\node[anchor=west] at (12*\dy,2.5*\dy) {\footnotesize $\mathbb{E}$};
	\node[anchor=west] at (12*\dy,.5*\dy) {\footnotesize $\mathbb{E} C_\Gamma \mathbb{E}$};
	\node[anchor=east] at (-2*\dy,8.5*\dy) {\footnotesize $C$};
	\node[anchor=east] at (-2*\dy,1*\dy) {\footnotesize $\ket{R}\bra{L}$};
	\node[anchor=east] at  (-2*\dy,5*\dy) {\footnotesize $\mathbb{E}$};
	\end{tikzpicture}
	}
 \caption{ 
 Contributions to the calculation of $S_{AB}$. Corners related by parity lead to the same contribution. }
\label{entropyregion}
\end{figure}

For the regions contributing to $S^{\text{KP}}_{\text{topo}}$, this leads to 
\begin{align}
S_B = S_C = & ( \bra{L} \mathbb{E}^{l_x-2} C  \mathbb{E}^{l_y-2} \ket{R} )^2
\, ,
\\
S_{AB} =&  \bra{L} \mathbb{E}^{l_x-2} C  \mathbb{E}^{ l_y- 1} C_\Gamma  \mathbb{E}^{l_x-1} C  \mathbb{E}^{l_y-2}  \ket{R}
 \bra{L} \mathbb{E}^{2 l_x-2} C  \mathbb{E}^{2 l_y-2} \ket{R}
  \, ,
\\
S_{BC} = &  \bra{L} \mathbb{E}^{2 l_x-2} C  \mathbb{E}^{ l_y-2} \ket{R} \bra{L} \mathbb{E}^{2 l_x-2} \ket{R_\Gamma} 
 \bra{L_\Gamma} \mathbb{E}^{ l_y -2} \ket{R} \bra{L} \mathbb{E}^{ l_x-2} C  \mathbb{E}^{2 l_y-2} \ket{R}
 \, ,
\\
S_{ABC} =&  ( \bra{L} \mathbb{E}^{2 l_x-2} C  \mathbb{E}^{2 l_y-2} \ket{R} )^2
\end{align}
where we have assumed $l_x'=l_x$ and $l_y'=l_y$ for simplicity, as $S_A$ and $S_{BC}$ cancel in this case.

\begin{figure}[t]
\centering
\begin{subfigure}[t]{0.49\linewidth}
\centering
 {{
 \includegraphics[scale=0.5]{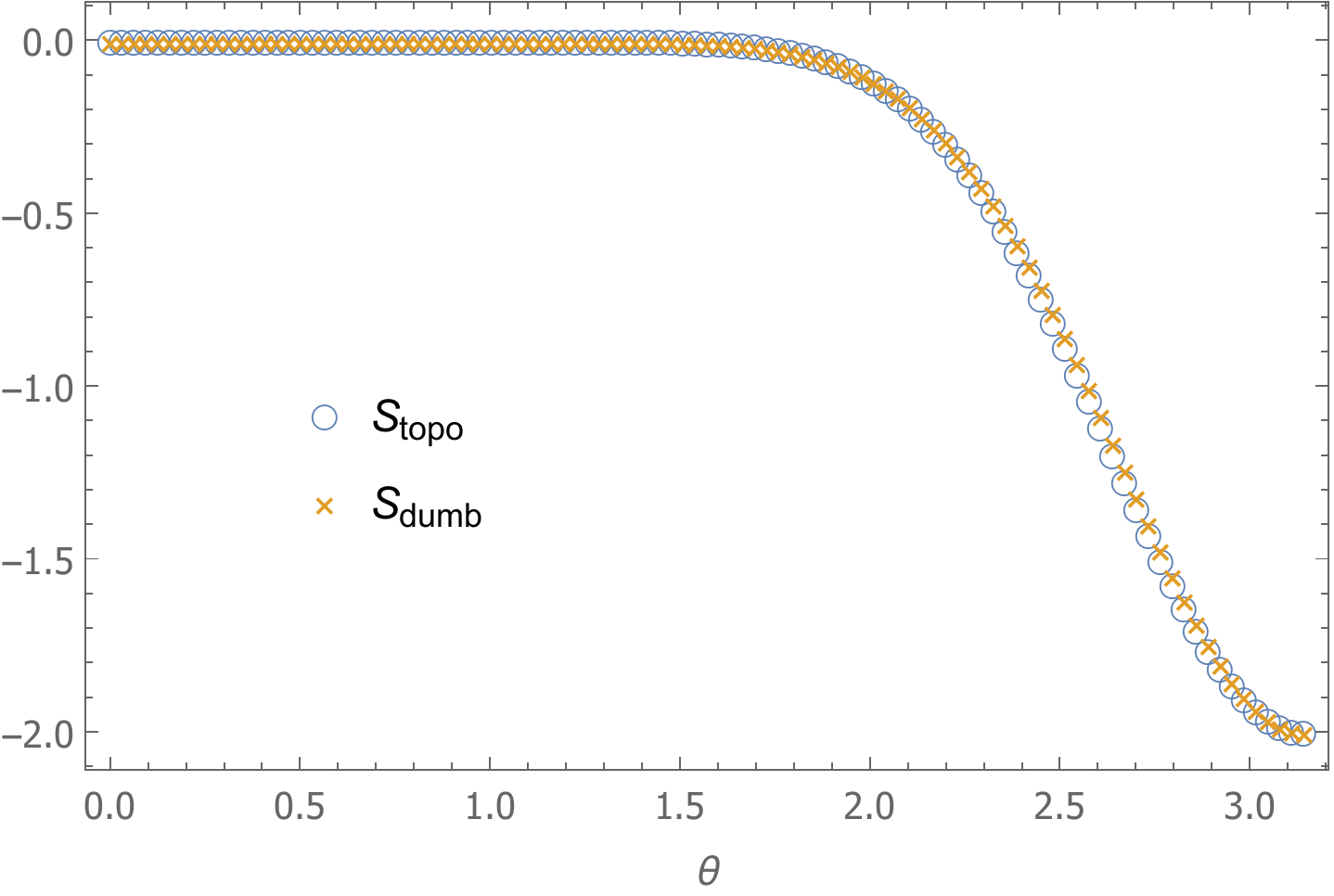} }} 
  \end{subfigure}
  \begin{subfigure}[t]{0.49\linewidth}
  \centering
 \includegraphics[scale=0.5]{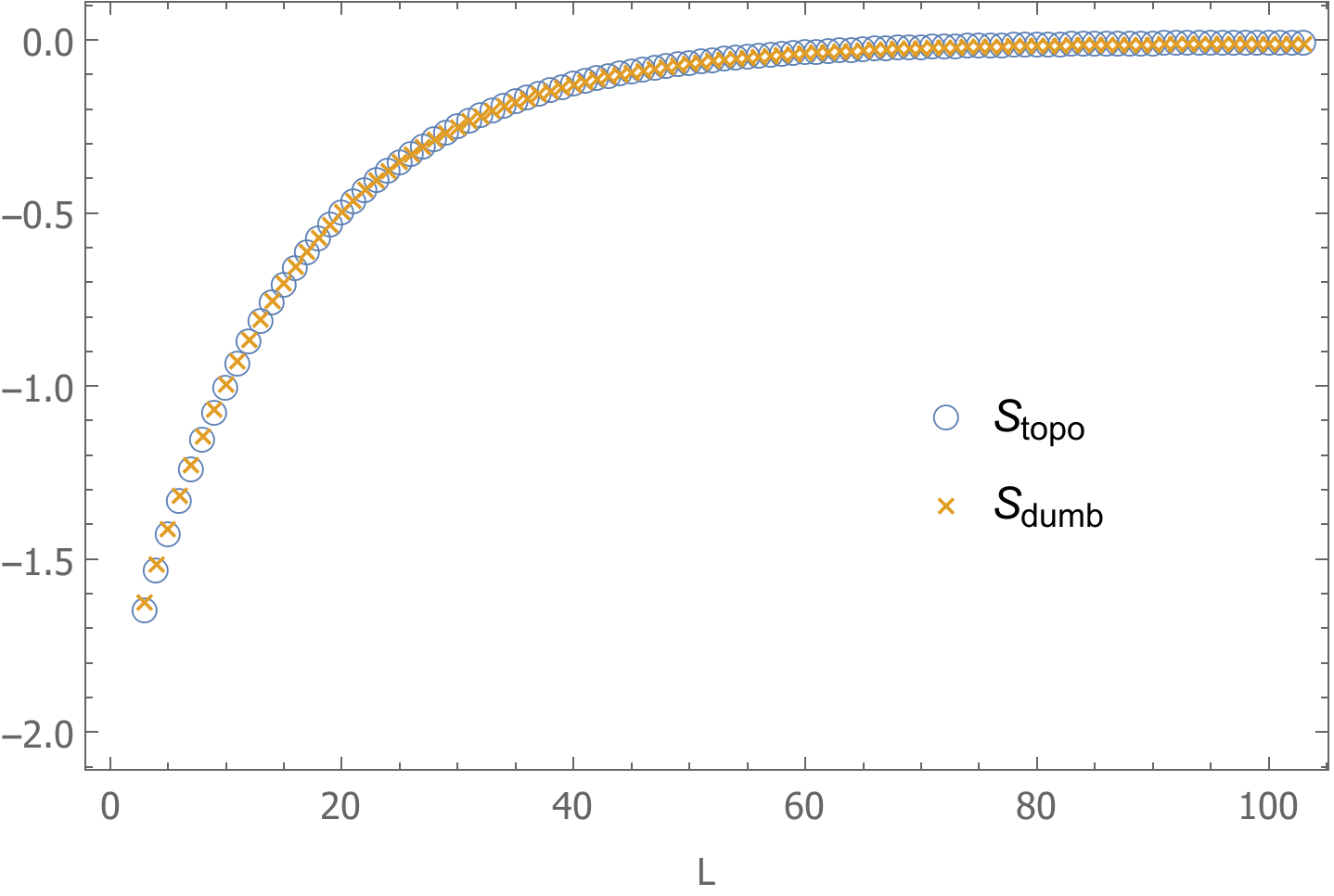}
  \end{subfigure}
 \caption{$S^{\text{KP}}_{\text{topo}},\, S_{\text{dumb}}$ for the deformed cluster state as functions of: $\theta$, for $l^{(')}_{x/y}=3$  (left), $L$, for $\theta=0.9 \pi$ (right).}
\label{TEEL}
\end{figure}

\begin{figure}[hb]
\center
\begin{subfigure}[t]{0.49\linewidth}
{{
 \includegraphics[scale=0.5]{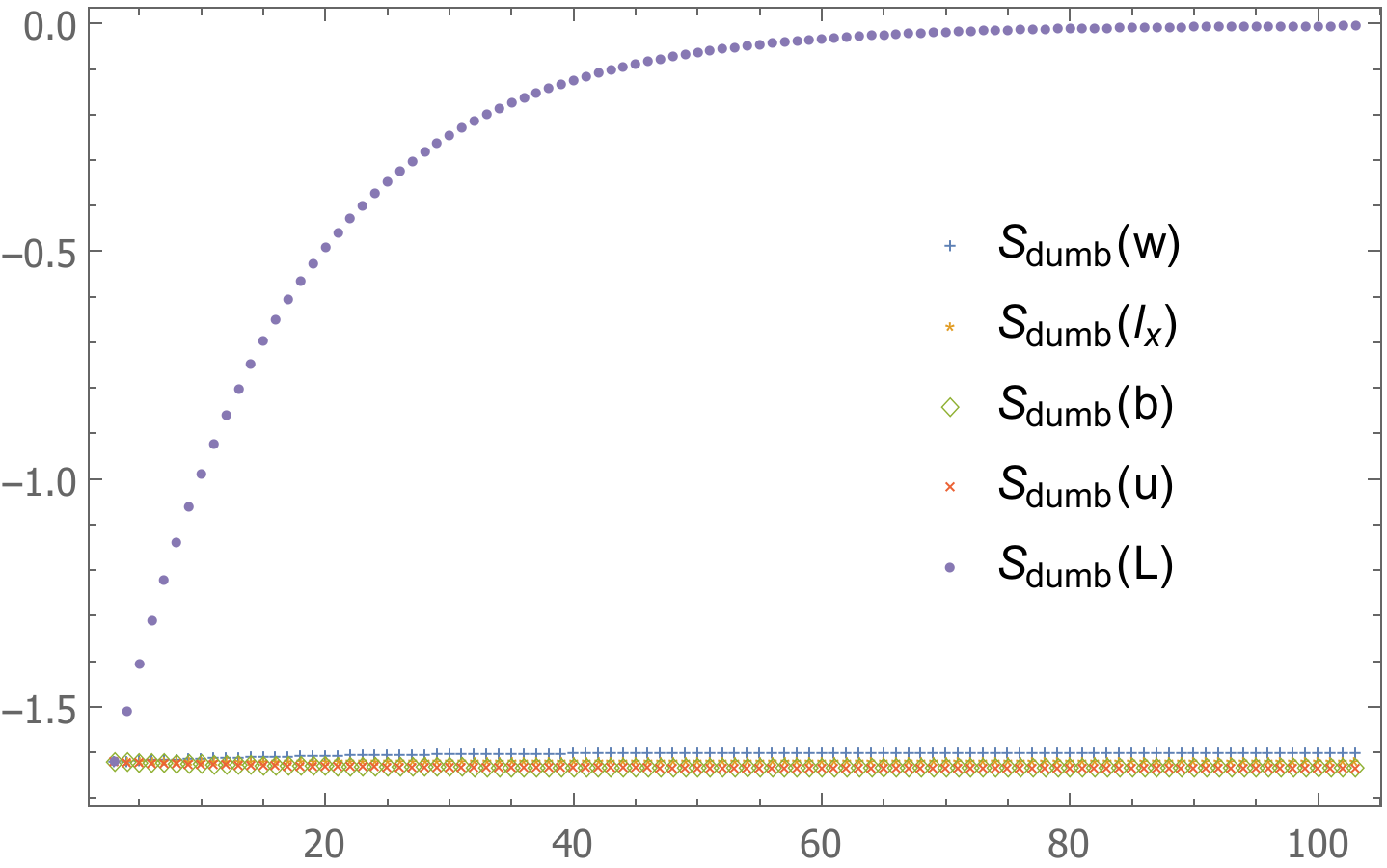} }}
\end{subfigure}
\begin{subfigure}[t]{0.49\linewidth}
{{
 \includegraphics[scale=0.5]{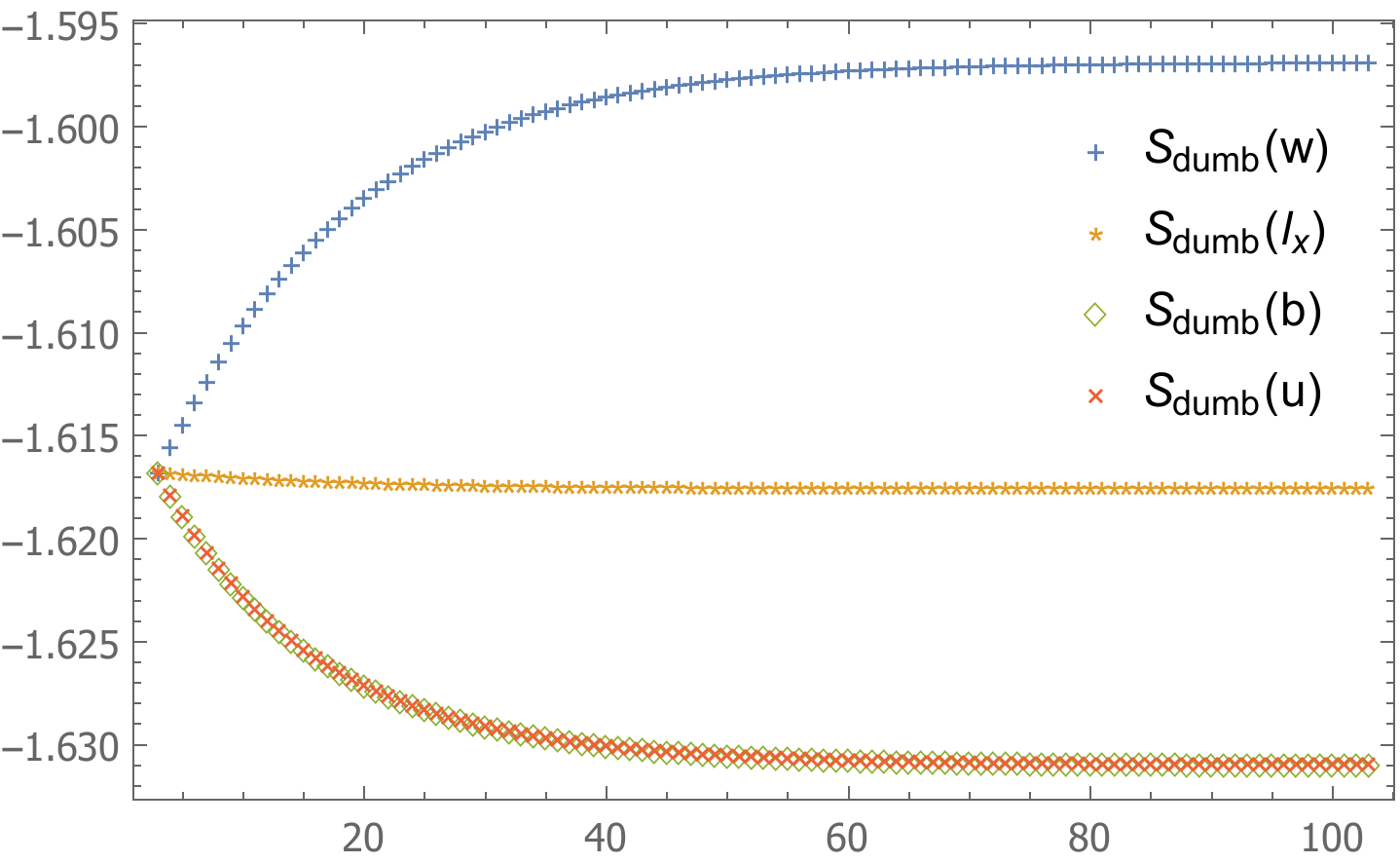} }}
\end{subfigure}
\caption{ (left) $S_{\text{dumb}}$ as different lengths are varied, for $\theta=0.9 \pi$. (right) More detailed plot of $S_{\text{dumb}}$. }
\label{Sdumballtheta}
\end{figure}

While the regions contributing to $S_{\text{dumb}}$  lead to 
\begin{align}
S_B =&  \bra{L} \mathbb{E}^{L-2} C  \mathbb{E}^{w-2} \ket{R} 
\, ,
\\
S_{AB} =&  \bra{L} \mathbb{E}^{l_x-2} C  \mathbb{E}^{ b- 1} C_\Gamma  \mathbb{E}^{L-1} C  \mathbb{E}^{w-2}  \ket{R}
\bra{L} \mathbb{E}^{L-2} \ket{R_\Gamma}  
  \bra{L_\Gamma} \mathbb{E}^{t-2} \ket{R}   \bra{L} \mathbb{E}^{l_x-2} C \mathbb{E}^{l_y-2}  \ket{R}
  \, ,
\\
S_{BC} = &    \bra{L} \mathbb{E}^{L-2} \ket{R_\Gamma}   \bra{L_\Gamma} \mathbb{E}^{t-2} \ket{R}   \bra{L} \mathbb{E}^{l_x-2} C \mathbb{E}^{l_y-2}  \ket{R}
  \bra{L} \mathbb{E}^{l_x-2} C  \mathbb{E}^{ t- 1} C_\Gamma  \mathbb{E}^{L-1} C  \mathbb{E}^{w-2}  \ket{R}
  \, ,
\\
S_{ABC} =& \bra{L} \mathbb{E}^{l_x-2} C  \mathbb{E}^{ b - 1} C_\Gamma  \mathbb{E}^{L-1}   \ket{R_\Gamma} 
\bra{L_\Gamma} \mathbb{E}^{b-2} \ket{R}  
  \bra{L} \mathbb{E}^{l_x -2} C \mathbb{E}^{l_y -2}  \ket{R} 
\bra{L} \mathbb{E}^{l_x-2} C  \mathbb{E}^{ u - 1} C_\Gamma  \mathbb{E}^{L-1}  \ket{R_\Gamma}
\nonumber \\
& \bra{L_\Gamma} \mathbb{E}^{t-2} \ket{R}   
 \bra{L} \mathbb{E}^{l_x-2} C \mathbb{E}^{l_y-2}  \ket{R}
 \, .
\end{align}
The results are shown in Figs.~\ref{TEEL},~\ref{Sdumballtheta}.

A nonzero value of $S_\text{dumb}$ is closely related to an infinite replica correlation length as defined in Ref.~\onlinecite{PhysRevB.94.075151}. In particular, consider a model with sites on the vertices of a square lattice and stabilizer generators on each plaquette. 
Assume there are subsystem symmetry generators given by tensor products of $X$ operators along rows or columns that are a single site wide. Further assume that each such rigid symmetry generator is not a product of local stabilizers with support on the relevant width one row (column). 
For such a model, the replica correlation length is infinite when calculated for a region consisting of an infinite strip (row or column) that is a single site wide. This is because one can use a pair of single site $Z$ operators, that each anti commute with the same $X$ line symmetry and are taken arbitrarily far apart, to construct a replica correlation function that is always equal to $1$. This analysis applies, in particular, to the cluster model considered in the body of the paper. More generally, it applies to any stabilizer model with $X$ subsystem symmetries and stabilizer generators that act nontrivially on all corners adjacent to a plaquette (technically we must further assume that no stabilizer in the model has a smaller support than these generators).

\end{document}